\documentclass[12pt]{article}
\usepackage{epsfig,here,rotate}
\def\gappeq{\mathrel{\rlap {\raise.5ex\hbox{$>$}}
{\lower.5ex\hbox{$\sim$}}}}

\def\lappeq{\mathrel{\rlap{\raise.5ex\hbox{$<$}}
{\lower.5ex\hbox{$\sim$}}}}

\begin{document}
\setcounter{footnote}{0}
\newcommand{\mycomm}[1]{\hfill\break
$\phantom{a}$\kern-3.5em{\tt===$>$ \bf #1}\hfill\break}
\newcommand{\mycommA}[1]{\hfill\break
$\phantom{a}$\kern-3.5em{\tt***$>$ \bf #1}\hfill\break}
\renewcommand{\thefootnote}{\fnsymbol{footnote}}
\newcommand{\ksl}{\mbox{$k$\hspace{-0.5em}\raisebox{0.1ex}{$/$}}}
\newcommand{\psl}{\mbox{$p$\hspace{-0.4em}\raisebox{0.1ex}{$/$}}}
\newcommand{\pbsl}{\mbox{$\bar{p}$\hspace{-0.4em}\raisebox{0.1ex}{$/$}}}
\newcommand{\qsl}{\mbox{$q$\hspace{-0.45em}\raisebox{0.1ex}{$/$}}}

\catcode`\@=11 
\def\lsim{\mathrel{\mathpalette\@versim<}}
\def\gsim{\mathrel{\mathpalette\@versim>}}
\def\@versim#1#2{\vcenter{\offinterlineskip
        \ialign{$\m@th#1\hfil##\hfil$\crcr#2\crcr\sim\crcr } }}
\catcode`\@=12 
\def\beq{\begin{equation}}
\def\eeq{\end{equation}}
\def\MSbar {\hbox{$\overline{\hbox{\tiny MS}}\,$}}
\def\eff{\hbox{\tiny eff}}
\def\res{\hbox{\tiny res}}
\def\FP{\hbox{\tiny FP}}
\def\PV{\hbox{\tiny PV}}
\def\IR{\hbox{\tiny IR}}
\def\UV{\hbox{\tiny UV}}
\def\ECH{\hbox{\tiny ECH}}
\def\NP{\hbox{\tiny NP}}
\def\APT{\hbox{\tiny APT}}
\def\QCD{\hbox{\tiny QCD}}
\def\CMW{\hbox{\tiny CMW}}
\def\SDG{\hbox{\tiny SDG}}
\def\spins{\hbox{\tiny spins}}
\def\SDG{\hbox{\tiny SDG}}
\def\pinch{\hbox{\tiny pinch}}
\def\brem{\hbox{\tiny brem}}
\def\V{\hbox{\tiny V}}
\def\BLM{\hbox{\tiny BLM}}
\def\NLO{\hbox{\tiny NLO}}
\def\PT{\hbox{\tiny PT}}
\def\PA{\hbox{\tiny PA}}
\def\1loop{\hbox{\tiny 1-loop}}
\def\2loop{\hbox{\tiny 2-loop}}
\def\mysim{\kern -.1667em\lower0.8ex\hbox{$\tilde{\phantom{a}}$}}
\def\a{\bar{a}}

\begin{titlepage}
\begin{flushright}
{\small CERN-TH/2001-229}\\
{\small August, 2001}

\end{flushright}
\vspace{.13in}

\begin{center}
{\Large {\bf Dressed gluon exponentiation}}\footnote{Research
supported in part by the EC
program ``Training and Mobility of Researchers'', Network
``QCD and Particle Structure'', contract ERBFMRXCT980194.}

\vspace{.4in}

{\bf Einan Gardi}

\vspace{0.25in}

TH Division, CERN, CH-1211 Geneva 23, Switzerland\\

\vspace{.4in}

\end{center}
\noindent  {\bf Abstract:} Perturbative and non-perturbative
aspects of differential cross-sections close to a kinematic
threshold are studied applying ``dressed gluon exponentiation'' (DGE).
The factorization property of soft and collinear gluon radiation
is demonstrated using the light-cone axial gauge: it is shown that
the singular part of the squared matrix element for the emission
of an {\em off-shell} gluon off a nearly on-shell quark is
universal. We derive a generalized splitting function that
describes the emission probability and show how Sudakov logs
emerge from the phase-space boundary where the gluon transverse
momentum vanishes. Both soft and collinear logs associated with a
single dressed gluon are computed through a single integral over
the running-coupling to any logarithmic accuracy. The result then
serves as the kernel for exponentiation. The divergence of the
perturbative series in the exponent indicates specific
non-perturbative corrections. 
We identify two classes of
observables according to whether the radiation is from an
initial-state quark, as in the Drell--Yan process, or a final-state
quark, forming a jet with a constrained invariant mass, as in fragmentation 
functions, event-shape variables and deep inelastic structure 
functions. \vspace{.25in}
\end{titlepage}

\section{Introduction}

Many observables in QCD receive large corrections from soft and
collinear gluon radiation.  Contrary to virtual corrections, real
gluon emission has a restricted phase space that depends on the
kinematics.  In particular, close to a kinematic threshold, a
differential cross-section is dictated by soft and collinear gluon
radiation. Classical examples are event-shape variables close to
the two-jet limit, and fragmentation functions, deep inelastic
scattering structure functions and Drell--Yan production for $x
\longrightarrow 1$.

Differential cross-sections close to threshold have been studied
extensively in perturbation theory,  where the phenomenon is
reflected in Sudakov logs $L=\ln 1/(1-x)$. The presence of large
logs in the perturbative coefficients limits the applicability of
fixed-order  calculations. It is a universally
acknowledged conjecture that perturbation theory can be improved  by
resumming the leading logs to all orders in the
coupling~\cite{Basics}--\cite{Contopanagos:1997nh}.

The resummation of Sudakov logs is based on the general property
of factorization. Factorization of soft photons (gluons)  goes
back to the Low theorem~\cite{Low,Burnett:1968km},  which states
that the leading, ${\cal{O}} (1/w)$ ($w$ is the photon energy),
and next-to-leading, ${\cal{O}} (w^0)$, terms in the radiative
amplitude in a generic process can be expressed in terms of the
radiationless amplitude. As a consequence, in the soft
approximation the amplitude can be written as a product of a
universal bremsstrahlung factor times the  amplitude that
describes the hard process. Under certain conditions, the
bremsstrahlung factorization theorem can be extended \cite{Gribov} to include
hard collinear radiation with small transverse 
momentum~$k_t$~(see also \cite{Basics}--\cite{Collins:1988ig} and~\cite{DelDuca:1990gz,Chaichian:1995kq}).

An important consequence of factorization is that the log-enhanced
part of the perturbative expansion of differential cross-sections
{\em exponentiates}. This means that a leading order  calculation
of the single parton branching probability  is sufficient to
generate the leading contributions to any order in perturbation
theory.

In single-scale observables renormalization group and physical
considerations like~\cite{BLM} allow one to choose the
renormalization scale~$\mu^2$ such that unnecessarily large logs
are avoided. In differential cross-sections there are several
physical scales, e.g. the hard scale $Q^2$ and $Q^2(1-x)$, so
large logs are inherent. The choice of the renormalization point
as the hardest scale $\mu^2\simeq Q^2$ will make the coupling
small. However, for $x\longrightarrow 1$ the cross-section is
determined by multiple soft and collinear emission with typical
momenta that are much smaller than $Q$. Using $\alpha_s(Q^2)$ as
an expansion parameter will force the coefficients to be large,
reflecting the presence of the lower scale~$Q^2(1-x)$. As a
consequence the resummation of the differential cross-section is
strictly restricted to the range $Q^2(1-x)>\bar{\Lambda}^2$ where
$\bar{\Lambda}$ is the scale  at which the {\em running coupling}
diverges. In addition, within the perturbative domain, perturbative
and non-perturbative corrections are large.

For fragmentation functions and deep inelastic structure functions, 
which are both governed by collinear emission, 
the perturbative domain is $Q^2(1-x)>\bar{\Lambda}^2$. In event-shape distribution, and in
Drell--Yan production close to threshold, large angle soft emission
enforces a more stringent restriction, e.g. $Q(1-T)>\bar{\Lambda}$
in the case of the thrust ($T$). In terms of the coupling, 
the largest possible domain of applicability of Sudakov resummation is
therefore $AL<\xi_{\max}$, where $A\equiv \beta_0\alpha_s(Q^2)/\pi$
and $\xi_{max}=1$ or $1/2$ in the two cases mentioned.

It is well known that perturbative coefficients in QCD 
increase factorially at large orders. The dominant source of
divergence are diagrams that are related to the running coupling,
the so-called ``renormalons''. Due to renormalons, the
applicability of fixed-order perturbation theory may be more
restricted than a priori expected. The problem can be addressed in
perturbation theory by the resummation of the running coupling
effects, using the analogy with the Abelian
theory~\cite{BLM}--\cite{beneke}.

It is important to realize that a perturbative treatment of
running coupling effects is not sufficient: infrared renormalons
make the resummation procedure ambiguous at power accuracy. For
some inclusive observables, such as the total cross-section in
$e^+e^-$ annihilation, power corrections are small~$\sim 1/Q^{4}$.
This can be established by an operator product expansion or by the
analysis of infrared renormalon
ambiguity~\cite{Mueller,Zakharov,beneke}.  On the other hand, in
differential cross-sections, the leading corrections are
determined by the lower scale, i.e. $\sim 1/(Q^{2}(1-x))^n$ (or,
for event-shape variables such as the thrust, $1/(Q(1-T))^n$).
Consequently also the power-correction expansion is expected to
break down for $Q^2(1-x)\gsim \bar{\Lambda}^2$ (or for
$Q(1-T)\gsim \bar{\Lambda}$ in the thrust case). The positive side
is that the factorization property holds beyond the
perturbative level. It is natural to expect that like the perturbative series also 
the power-correction series exponentiates close to the kinematic threshold~\cite{Shape_function1}--\cite{DGE}.

The impact of renormalons on differential cross-sections is
both interesting in principle and important in practice. The
analysis of the thrust distribution $e^+e^-$ annihilation in
\cite{DGE} showed that due to renormalons,  sub-leading  logs are
factorially enhanced with respect to the leading logs. This
observation is completely general: this is the way in which
sensitivity to lower scales is reflected in the perturbative
coefficients. The immediate consequence is that resummation based
on a fixed logarithmic accuracy has a limited range of
validity, $LA\ll \xi_{\max}$. A quantitative treatment in a wider
range (still $LA <\xi_{\max}$) requires to sum all the factorially
enhanced sub-leading logs and include the associated
power corrections in the exponent.  This is achieved by Dressed
Gluon Exponentiation~(DGE)~\cite{DGE}.

DGE is designed to deal with both Sudakov logs and renormalons. It
is based on exponentiating the entire perturbative series (of
log-enhanced terms) that describes the single gluon emission close
to the threshold. At a difference from the standard parton cascade
approach (see~e.g.~\cite{CT_DY,CTTW}), the single gluon emission
probability is calculated to any order in perturbation theory (in
the large-$\beta_0$ limit) and to any logarithmic accuracy,  thus
incorporating the factorial enhancement of sub-leading logs. In
addition to the improved perturbative treatment, by identifying
the divergence of the exponent one can introduce parametrization
of the relevant non-perturbative corrections.

A good example where renormalon analysis is crucial in identifying
the leading power corrections is the case of Drell--Yan
production~\cite{Shape_function1}, \cite{CS_DY}--\cite{AZ_KLN}. Here,
kinematic considerations suggest that as in the thrust
distribution, soft gluon emission at large angles will result in a
$1/Q(1-x)$ correction. Renormalon analysis~\cite{BB_DY,AZ_KLN}
shows that the leading correction is in fact $\sim 1/Q^{2}(1-x)^2$.
This difference between the case of event-shape variables and that
of Drell--Yan is very intriguing and we shall return to discuss it
below.

In both \cite{BB_DY} and \cite{DGE}, the starting point for the
perturbative calculation was that of the dispersive approach in
renormalon calculus~\cite{BB,DMW} (see also~\cite{beneke,Average_thrust}):
evaluation of the single dressed gluon (SDG) cross-section based
on the exact matrix element and the exact {\em off-shell} gluon
phase space. On the other hand, the factorization property and the
resulting Sudakov log resummation
methodology (see e.g.~\cite{CT_DY,Collins:1988ig,Contopanagos:1997nh,CTTW})
are based on keeping the gluon {\rm on-shell} and  making specific
kinematic approximations that single out the singular parts of
the gluon emission matrix element. Although integrals over the
running coupling have a central role in both cases, the
scale of the coupling in the former is the gluon virtuality, 
whereas in the latter it is its transverse momentum. Thus at first sight, the two
methodologies seem mutually exclusive. In this paper we show how
they can be bridged over. We identify the appropriate
approximation required to single out the singular part of the
matrix element in the case of an off-shell gluon emission. Next,
we show that the factorization property applies to off-shell
gluons in the same way it applies to on-shell gluons. This
allows us to fully exploit the power of factorization: the entire
renormalon sum exponentiates.

The calculation of the exponent is, of course, approximate.
However, instead of working at a fixed logarithmic accuracy, our
approximation is based on the Abelian large-$\beta_0$ limit. The
usual reason to consider this limit is that it identifies the
dependence on the scale of the running coupling and thus also the
infrared sensitivity. In the context of exponentiation there is an
additional motivation: the large-$\beta_0$ limit distinguishes
single gluon emission from multiple emission. Since the
exponentiation kernel is primarily associated with a single gluon
emission, it is natural to evaluate it using this limit. The first
step beyond the large-$\beta_0$ limit is taken in an unambiguous
way by identifying the running coupling  with the ``gluon
bremsstrahlung'' effective charge~\cite{CMW,DGE}. Using this
coupling and the two-loop renormalization group equation is sufficient
to guarantee that the DGE resums {\em all} the logs (i.e. not only
the large-$\beta_0$ logs) up to next-to-leading logarithmic
accuracy (NLL). Contrary to a fixed logarithmic accuracy
resummation, DGE is free of renormalization-scale dependence.
Clearly, some residual scale dependence appears beyond the
logarithmic approximation upon matching the exponentiated result
with fixed-order calculations~\cite{DGE}. However, in the
threshold region, when the logarithmically enhanced terms
dominate, this effect is negligible.

The logarithmically enhanced cross-section can be calculated from
the singular terms in the squared matrix element. Using the
light-cone axial gauge we demonstrate that as far as the singular
terms are concerned, the emission probability is
process-independent. The squared matrix element factorizes into a
hard part times a function that depends only on the gluon
momentum. The latter is interpreted as a generalized splitting
function. The factorization of the cross-section depends on the
phase space. Here we consider the case where all energetic
particles move in two light-cone directions. Such light-cone
kinematics arises in different cases: in $e^+e^-$ these are the
two opposite directions of the quark jets, in Drell--Yan production
near $x=1$ these are the directions of the two incoming quarks,
and in deep inelastic scattering near $x_{\rm Bj}=1$ one light-cone
direction is the direction of the incoming quark, the other
that of the outgoing quark. In each of these cases we show how the
singularity of the matrix element translates into Sudakov logs
through the phase-space integration. It is crucial that the logs
emerge only from the limit in which the gluon transverse momentum
vanishes (it can be soft or collinear). As a consequence, the
relevant terms in the characteristic function can be easily
computed. The resulting characteristic function is integrated
with the running coupling yielding a resummed
cross-section that contains {\em all} the logarithmic terms
associated with a single dressed gluon. This is the kernel for
DGE.

The outline of this paper is as follows. In section 2 we use the
light-cone gauge to analyse the generic scenario of an off-shell
gluon bremsstrahlung in light-cone kinematics. We identify the
approximation needed to isolate the singular parts of the squared
matrix element and show that in this approximation the radiative
squared matrix element can be written in a factorized form in
which a universal bremsstrahlung factor multiplies the
non-radiative squared matrix element. In section 3 we turn to
discuss factorization at the level of the cross-section.
Considering first the case of final-state radiation in $e^+e^-$
annihilation we derive a generalized splitting function formula
for the case of an off-shell gluon. In section 4 we illustrate the
advantage of the axial gauge in this context. Using the case of
the quark fragmentation function as an example, sections 5 and 6
describe the two steps in promoting the leading order calculation
to a resummed one: in section 5 we show how phase-space
integration can be performed within the approximation considered
to obtain the relevant part of the SDG characteristic function and
then use the dispersive approach to dress the gluon through an
integral over the running coupling. In section 6 we show how
multiple emission can be taken into account by exponentiation of
the SDG cross-section. Sections 7 though 9 summarize the
application of the method to other observables: jet mass (or 
thrust) in $e^+e^-$, coefficient functions in deep inelastic
scattering and Drell--Yan production. Common features and
differences between the different examples are emphasized. Section
10 contains some concluding remarks.

\section{Factorization of the squared matrix element}

Sudakov logs appear because of soft or collinear gluon radiation off
nearly on-shell partons. The emitting parton can be a gluon or a
quark. In the standard treatment of parton cascades, each possible
branching is assigned a probability, which is a function of the
longitudinal momentum fraction. Here we develop a similar
probabilistic formalism for the case of an off-shell gluon
emission off a quark. Gluon splitting will be taken into account
in a different way, through an integral over the running coupling.

In a generic QCD process, consider gluon emission off an outgoing
quark, assuming that the quark is on-shell after the emission
$p^2=m_q^2\simeq 0$, while the gluon possesses a time-like
virtuality $k^2=m^2$.
\begin{figure}[t]
\begin{center}
\mbox{\kern-0.5cm\epsfig{file=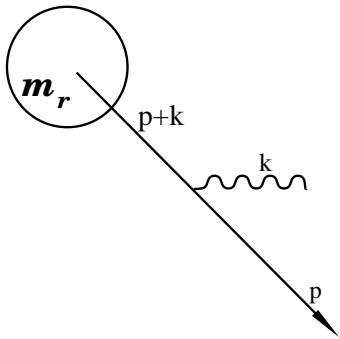,width=4.5truecm,angle=0}}
\end{center}
\caption{Gluon bremsstrahlung off an outgoing quark}
\label{log_func}
\end{figure}
The amplitude for the emission off this particular quark is
\begin{equation}
\label{Amp}
{\cal M}=g_s\,t^a\,{\epsilon^{\lambda}_{\mu}}^*\,\frac{1}{(k+p)^2}\,\bar{u}^{(s)}(p)
\gamma^\mu(\psl+\ksl){\cal M}_r,
\end{equation}
where $t^a$ is a colour matrix, ${\epsilon^{\lambda}_{\mu}}^*$ is
the gluon polarization vector and ${\cal M}_r$ represents the rest
of the process. The enhancement of soft or collinear emission is
associated with the singularity of the quark propagator
$1/(k+p)^2$, which is only partially compensated by the terms in
the numerator.  Factorization of the cross-section is based on the
fact that the singularity emerges just from this diagram. The
squared matrix element corresponding to (\ref{Amp}) summed over
the gluon and quark spins is
\begin{eqnarray}
\label{M2}
\sum_{\mbox {\rm {\spins}}} {\cal{M}}{\cal{M}}^\dagger = \frac{C_F g_s^2}{(k+p)^4}\,
\left( \sum_{\lambda}{\epsilon_\nu^{\lambda}}^*\epsilon_\mu^{\lambda}\right)
{\rm Tr}\left\{(\psl+\ksl)\gamma^\nu
\psl \gamma^\mu (\psl+\ksl){\cal M}_r{{\cal M}_r}^{\dagger}\right\}
\end{eqnarray}
where $\sum_{s} u^{(s)}(p)\bar{u}^{(s)}(p)$ was replaced by
$\psl$. Since the gluon attaches also to other particles with an
amplitude $\bar{{\cal M}}$, the full calculation of the
cross-section should be based on \hbox{$({\cal M}+\bar{{\cal
M}})({\cal M}+\bar{{\cal M}})^\dagger$} rather than on (\ref{M2}).
However, the explicit terms in (\ref{M2}) capture the {\em entire}
singularity {\em if }  interference terms of the form ${\cal
M}{\bar{{\cal M}}}^\dagger$ are non-singular. This depends on the
gauge choice
$\sum_{\lambda}{\epsilon_\nu^{\lambda}}^*\epsilon_\mu^{\lambda}\equiv
d_{\mu \nu}$.

Defining two light-like vectors by $p=(p_{+},0,0)$ and
$\bar{p}=(0,\bar{p}_{-},0)$, where \hbox{$p^2=\bar{p}^2=0$}, we
choose the light-cone axial gauge where $A_+=0$, which is a
physical gauge: \hbox{$g^{\mu \nu}d_{\mu \nu}=-2$}. The gluon
propagator is given by
\begin{eqnarray}
\label{propagator}
d_{\mu \nu}=-g_{\mu \nu}+\frac{k_{\mu}\bar{p}_\nu+\bar{p}_{\mu}k_\nu}{k\bar{p}}.
\end{eqnarray}
Let us consider now light-cone kinematics. Particles moving in the
``$-$'' direction couple to the gauge field through $J_-A_+$,
where $J_-$ is an effective current. Since in this gauge
\hbox{$A_+$} vanishes, these particles do not interact with the
gauge field. Therefore, if we assume that the other coloured
particles move close enough to the ``$-$'' direction, it is
sufficient to consider the amplitude in~(\ref{Amp}). Put
differently, in this gauge (\ref{M2}) captures the entire
singularity and the interference terms ${\cal M}{\bar{{\cal
M}}}^\dagger$ are non-singular. This point will be demonstrated
explicitly below.

Note that the assumption we made here is rather strong: the
invariant mass of the particles moving in the ``$-$'' direction
{\em must be negligible}. This assumption is relevant in many
applications including event-shape variables close to the two-jet
limit, and fragmentation functions, deep inelastic scattering
structure functions and Drell--Yan production for $x
\longrightarrow 1$. On the other hand, this assumption is
inadequate for a generic jet observable in hadron--hadron
collisions where both initial- and final-state radiation are
important, nor for the case of three jets in $e^+e^-$
annihilation.

Let us now calculate the squared matrix element (\ref{M2}) in the light-cone gauge.
The traces corresponding to the two terms in the propagator~(\ref{propagator}) are
\begin{eqnarray}
r_1&\equiv&2\,{\rm Tr} \left\{ (\psl+\ksl) \psl (\psl+\ksl) {\cal M}_r{{\cal M}_r}^{\dagger}\right\} \\ \nonumber
&=&-2m^2\,{\rm Tr} \left\{\psl {\cal M}_r{{\cal M}_r}^{\dagger}\right\}
+4pk\,{\rm Tr} \left\{  \ksl {\cal M}_r{{\cal M}_r}^{\dagger}\right\} \\ \nonumber
r_2&\equiv&\frac{1}{k\bar{p}}\left[{\rm Tr}
\left\{ (\psl+\ksl) \pbsl\psl\ksl (\psl+\ksl){\cal M}_r{{\cal M}_r}^{\dagger}\right\}+\,
{\rm Tr}
\left\{ (\psl+\ksl) \ksl\psl\pbsl (\psl+\ksl){\cal M}_r{{\cal M}_r}^{\dagger}\right\}\right] \\ \nonumber
&=&\frac{(k+p)^2}{k\bar{p}}\left[(4p\bar{p}+2k\bar{p})\,{\rm Tr} \left\{\psl {\cal M}_r{{\cal M}_r}^{\dagger}\right\}
+2p\bar{p}\,{\rm Tr} \left\{  \ksl {\cal M}_r{{\cal M}_r}^{\dagger}\right\}
-2pk\,{\rm Tr} \left\{  \pbsl {\cal M}_r{{\cal M}_r}^{\dagger}\right\}  \right].
\end{eqnarray}
Collecting the terms $r\equiv r_1+r_2$ one obtains
\begin{eqnarray}
r &=&{\rm Tr}\left\{\psl {\cal M}_r{{\cal M}_r}^{\dagger}\right\}\left[-2m^2+\frac{(k+p)^2}{k\bar{p}}(4p\bar{p}+2k\bar{p})
\right] \nonumber \\ &+&
{\rm Tr} \left\{ \ksl {\cal M}_r{{\cal M}_r}^{\dagger}\right\}\left[4pk+2p\bar{p}\frac{(k+p)^2}{k\bar{p}}
\right]\,+\,
{\rm Tr} \left\{ \pbsl {\cal M}_r{{\cal M}_r}^{\dagger}\right\}\left[\frac{-2pk (k+p)^2}{k\bar{p}}
\right].
\end{eqnarray}

Introducing the Sudakov decomposition of $k=(k_{+},k_{-},k_t)=\beta p+\alpha \bar{p}+k_t$, and
defining the following dimensionless parameters
\begin{eqnarray}
\label{lc_par}
\beta&=&k_{+}/p_{+}=2k\bar{p}/2p\bar{p}\nonumber \\
\alpha&=&k_{-}/\bar{p}_{-}=2kp/2p\bar{p}\\ \nonumber
\gamma&\equiv& k_t^2/2p\bar{p} \\ \nonumber
\lambda&=&m^2/2p\bar{p}=(2k_{+}k_{-}-k_t^2)/2p\bar{p}=\alpha\beta-\gamma
\end{eqnarray}
the trace becomes
\begin{eqnarray}
\label{t_lc}
r&=&2p\bar{p} \,{\rm Tr}\left\{\psl {\cal M}_r{{\cal M}_r}^{\dagger}\right\}
\left[-2\lambda+2\frac{\alpha+\lambda}{\beta}(2+\beta)\right] \nonumber\\
&+& 2p\bar{p} \,{\rm Tr} \left\{ \ksl {\cal M}_r{{\cal M}_r}^{\dagger}\right\}\left[2\alpha+2\frac{\alpha+\lambda}{\beta}
\right]+
2p\bar{p} \,{\rm Tr} \left\{ \pbsl {\cal M}_r{{\cal M}_r}^{\dagger}\right\}
\left[-2\alpha\frac{\alpha+\lambda}{\beta}\right].
\end{eqnarray}

Sudakov logs in the differential cross-section are associated with the region where the
denominator of the quark propagator $(p+k)^2\,=\,2pk+k^2\,=\,2p\bar{p}\,(\alpha+\lambda)$ is small.
Therefore, from now on we assume $\alpha+\lambda\ll 1$. Since both $\alpha$ and $\lambda$ are positive,
{\em both} must be small. The gluon on-shell condition $\alpha\beta=\gamma+\lambda$ then implies that
the transverse momentum  fraction $\gamma$ is small, at least with respect to $\beta$.
We are therefore interested in the contribution from the region where $\alpha$, $\lambda$ and
$\gamma$ are all small, but no specific hierarchy between them is imposed.
This approximation is adequate to treat both the soft and the collinear regions:
$\beta$ is small in the soft limit, but it is of order $1$ or larger in the collinear (non-soft) limit.

In this approximation one can simplify (\ref{t_lc}) by
\begin{itemize}
\item{}
replacing ${\rm Tr} \left\{ \ksl {\cal M}_r{{\cal M}_r}^{\dagger}\right\}$ by
$\beta {\rm Tr} \left\{ \psl {\cal M}_r{{\cal M}_r}^{\dagger}\right\}$. The other components of $k$ have additional
suppression by one of the small parameters.
\item{}
ignoring the terms proportional to
${\rm Tr} \left\{ \pbsl {\cal M}_r{{\cal M}_r}^{\dagger}\right\}$. 
Taking $\bar{p}_{-}$ of the order of $p_{+}$,
these are explicitly suppressed by $\alpha$ or $\lambda$, with respect to the relevant terms.
\end{itemize}
Then the following expression for the trace $r$ is obtained,
\begin{equation}
\label{trace}
r=\,2\,\left[
{-\lambda}\,(1+\beta)+({\alpha+\lambda})\,\frac{\beta^2+2\beta+2}{\beta}
\right]\,2p\bar{p}
\,{\rm Tr} \left\{ \psl {\cal M}_r{{\cal M}_r}^{\dagger}\right\},
\end{equation}
and, finally, the squared matrix element (\ref{M2}) is
\begin{eqnarray}
\label{M22}
 \sum_{\mbox {\rm {\spins}}} {\cal{M}}{\cal{M}}^\dagger = 2C_F g_s^2
\,\left[
\frac{-\lambda}{({\alpha+\lambda})^2}\,(1+\beta)+\frac{1}{\alpha+\lambda}\,\frac{\beta^2+2\beta+2}{\beta}
\right]\,\frac{{\rm Tr} \left\{ \psl {\cal M}_r{{\cal M}_r}^{\dagger}\right\}}{2p\bar{p}}.
\end{eqnarray}
This demonstrates that as far as the $\alpha+\lambda\longrightarrow 0$ singular terms are concerned,
the squared matrix element is factorized
into a {\em process-independent} factor, corresponding to the emission of an off-shell gluon, times
${\rm Tr} \left\{ \psl {\cal M}_r{{\cal M}_r}^{\dagger}\right\}$, which is the expression
for the squared matrix element in case of no gluon emission. The latter is, of course, process-dependent.

In the case of gluon radiation off an incoming quark, similar considerations yield
\begin{eqnarray}
\label{M2_DY}
\frac12 \sum_{\mbox {\rm {\spins}}} {\cal{M}}{\cal{M}}^\dagger = 2 C_F g_s^2
\,\left[
\frac{-\lambda}{({\alpha-\lambda})^2}\,(1-\beta)+\frac{1}{\alpha-\lambda}\,\frac{\beta^2-2\beta+2}{\beta}
\right]\,\frac{\frac12 \,{\rm Tr} \left\{ \psl {\cal M}_r{{\cal M}_r}^{\dagger}\right\}}{2p\bar{p}},
\end{eqnarray}
where the factor $1/2$ accounts for averaging over the quark spin.
This expression can be obtained directly from (\ref{M22}) by
taking $\alpha\longrightarrow -\alpha$ and $\beta\longrightarrow
-\beta$, corresponding to inverting the momenta $p$ and $\bar{p}$.
Note that in (\ref{M2_DY}), as in (\ref{M22}), the Sudakov
parameters $\alpha$, $\beta$, $\gamma$ and $\lambda$ are all
positive.

\section{Gluon emission probability}

To show factorization of the physical cross-section,
one has to consider the phase space.
Suppose that the external particle momenta are denoted by $p^{i}$ and $p^{f}$
for incoming and outgoing particles, respectively. The cross-section for $p^{i}_A+p^{i}_B
\longrightarrow \sum_n p^f_n+k$~is
\begin{eqnarray}
\label{cs}
\sigma= &&\!\!\!\!\!\frac{1}{4E_AE_B\left\vert v_A-v_B\right\vert} \,\prod_n
\int \frac{d^3\vec{p}^{f}_n}{(2\pi)^3}\frac{1}{2E^{f}_n}\,\\\nonumber &&
\int \frac{d^4k}{(2\pi)^3}\delta\left(k^2-m^2\right)\,(2\pi)^4 \,\delta^{(4)}\left(p^{i}_A+p^{i}_B-\sum_n
p^{f}_n-k\right)\,\sum_{\mbox {\rm {\spins}}} {\cal{M}}{\cal{M}}^\dagger.
\end{eqnarray}
Because of momentum conservation, the gluon is correlated with the
other final state particles. Nevertheless, kinematic factorization
can be established for a simple enough phase space in the
appropriate Mellin or Laplace space. The details of factorization
and the resulting exponentiation formula depend on the observable
(see e.g. \cite{CT_DY,Contopanagos:1997nh,CTTW}).

As an example we shall consider here (and in sections 4 through 7)
the case of $e^+e^-$ annihilation into hadrons. It is convenient
to identify the vectors that define the ``$+$'' and ``$-$''
directions $p$ and $\bar{p}$ as the final momenta of the two
primary quarks. Working in the approximation of independent
emission, we can calculate the gluon emission probability as if
there are only three outgoing particles $p$, $\bar{p}$ and $k$.
Multiple emission results, in this approximation, in
straightforward exponentiation, as described in section~6. The
centre-of-mass energy squared is $q^2$, where $q\equiv
p^{i}_A+p^{i}_B$. Note that it is not given by $2p\bar{p}$: the
contribution of gluon momentum {\em cannot} be neglected in the
collinear limit.

Assuming that factorization occurs, we proceed by performing the integration
over $p$ and $\bar{p}$ in the trivial case where
${\rm Tr} \left\{ \psl {\cal M}_r{{\cal M}_r}^{\dagger}\right\}$ is
proportional to $2p\bar{p}$. The three-particle phase-space integral is then
\begin{eqnarray}
\label{3PS}
\int \frac{d^4p}{(2\pi)^3}\,\delta\left(p^2\right)
\frac{d^4\bar{{p}}}{(2\pi)^3}\,\delta\left(\bar{p}^2\right)
\frac{d^4k}{(2\pi)^3}\,\delta\left(k^2-m^2\right)(2\pi)^4\delta^4(q-p-\bar{p}-k)= \nonumber \\
\frac{q^2}{128\pi^3}\int
\,d\alpha d\beta\frac{1-\lambda}{(1+\alpha+\beta+\lambda)^3}\simeq \frac{q^2}{128\pi^3}\int
\,d\alpha d\beta\frac{1}{(1+\beta)^3},
\end{eqnarray}
while the two particle phase-space integral is
\begin{eqnarray}
\label{2PS}
\int \frac{d^4p}{(2\pi)^3}\,\delta\left(p^2\right)
\frac{d^4\bar{{p}}}{(2\pi)^3}\,\delta\left(\bar{p}^2\right)(2\pi)^4\delta^4(q-p-\bar{p})
=\frac{1}{8\pi}.
\end{eqnarray}
Starting with the squared matrix element (\ref{M22}) integrated over the three-particle phase space,
we now identify the gluon emission
probability as the factor multiplying the cross-section for no gluon emission (which is
${\rm Tr} \left\{ \psl {\cal M}_r{{\cal M}_r}^{\dagger}\right\}$ integrated over the two-particle
phase space). Changing variables from $\lambda$ to
\beq
\epsilon\equiv
\frac{m^2}{Q^2}=\frac{\lambda\,(2p\bar{p})}{Q^2}=\frac{\lambda}{1+\alpha+\beta+\lambda}
\simeq\frac{\lambda}{1+\beta}
\label{epsilon_lambda}
\eeq
 guarantees that the integration is performed for fixed $m^2$ and $Q^2$.
The gluon emission probability is given by 
\beq
{dP}\,=\,\frac{C_F \alpha_s}{2\pi}\,\left[
\frac{-\lambda}{(\alpha+\lambda)^2}\,(1+\beta)+\frac{1}{\alpha+\lambda}\,\frac{\beta^2+2\beta+2}{\beta}
\right]\,\frac{d\alpha\,d\beta \,d\lambda \,\delta(\lambda-\epsilon(1+\beta))}{(1+\beta)^3}.
\label{splitting_function}
\eeq
Equation~(\ref{splitting_function}) can be
interpretated as a generalized splitting function for the off-shell gluon case.
The standard splitting function can be obtained in the limit $\lambda=0$: substituting $z=1/(1+\beta)$ as
the longitudinal momentum fraction of the quark, one recovers the standard formula
\[
dP= \frac{C_F\alpha_s}{2\pi}\,\frac{ 1+z^2}{1-z}dz \,d\ln\alpha.\]
A further
assumption that~$\beta\ll 1$ leads to the double logarithmic approximation.

Like the standard on-shell splitting function, eq.~(\ref{splitting_function}) should be understood
as an evolution kernel, so that multiple branching is described by iteration.
This is achieved by DGE~\cite{DGE}.

\section{Gauge choice and interference}

Before proceeding with the application of eq.~(\ref{splitting_function}), it is worth while to have a
second look at the matrix element in order to appreciate the significance of the gauge choice.
While the following discussion is quite general, to be concrete we will keep on using the example of
$e^+e^-\longrightarrow^{\!\!\!\!\!\!\!\!^{ \gamma}} \,\,\,\, q\bar{q}g$.
Considering the hadronic tensor $H_{\nu\rho}$,
the rest of the amplitude in~(\ref{Amp}) is ${\cal M}_r= \gamma^{\nu}\,u(\bar{p})$,
so the process-dependent trace in~(\ref{M22}) becomes
\beq
\sum_{\rm {\spins}}{\rm Tr} \left\{ \psl {\cal M}_r{{\cal M}_r}^{\dagger}\right\}=
 {\rm Tr}\left\{\psl \gamma^\nu \pbsl \gamma^\rho\right\}.
\label{Tr_ee}
\eeq
If the angle with respect the electron beam is integrated over, as in the total fragmentation function case,
it is sufficient to project $H_{\nu\rho}$ on $-g^{\nu\rho}$.
Then one can replace ${\rm Tr}\left\{\psl \gamma^\nu \pbsl \gamma^\rho\right\}$ by $8p\bar{p}$.
The light-cone gauge
result for the squared matrix element, based on gluon attachment to the $p$ quark only,
is therefore
\begin{eqnarray}
\label{lc_AA}
 \sum_{\rm {\spins}}\left.{\cal M}{{\cal M}}^\dagger\right\vert_{{\rm light-cone}}
=  a_0  \left[
\frac{-\lambda}{({\alpha+\lambda})^2}\,(1+\beta)+\frac{1}{\alpha+\lambda}\,\frac{\beta^2+2\beta+2}{\beta}
\right],
\end{eqnarray}
where
$a_0\equiv 8C_F g_s^2$,
but since the gluon attaches also to the antiquark $\bar{p}$ with an amplitude
\begin{equation}
\label{Amp_bar}
\bar{{\cal M}}=g_s\,t^a\,{\epsilon^{\lambda}_{\mu}}^*\,\frac{1}{(k+\bar{p})^2}\,\bar{u}^{(s)}(p)
\gamma^\nu(\pbsl+\ksl)\gamma^\mu u^{(s)}(\bar{p}),
\end{equation}
the full squared matrix element is
\beq
\sum_{\rm {\spins}}\left[{\cal M}{{\cal M}}^\dagger+
{\cal M}{\bar{{\cal M}}}^\dagger+\bar{{\cal M}}{{\cal M}}^\dagger
+\bar{{\cal M}}{\bar{{\cal M}}}^\dagger\right]\,=\,a_0
\left[
\frac{\alpha\beta-\lambda}{(\alpha+\lambda)^2}
+2\frac{1+\alpha+\beta+2\lambda}{(\alpha+\lambda)(\beta+\lambda)}
+\frac{\beta\alpha-\lambda}{(\beta+\lambda)^2}
\right].
\label{full_me}
\eeq

This result is of course gauge-invariant. Let us now examine the
origin of the various terms, in particular the singular terms for
$\alpha+\lambda \longrightarrow 0$, in different gauges. Let us
compare the Feynman gauge and the light-cone gauge. In the Feynman
gauge, ${\cal M}{{\cal M}}^\dagger$ is the first term in
(\ref{full_me}) and the interference is responsible for middle
term there. Thus in this gauge only part of the singularity is
captured by a calculation (${\cal M}{{\cal M}}^\dagger$) that
ignores the possibility of gluon attachment to the rest of the
process. On the other hand in the light-cone gauge the {\em
entire} $\alpha+\lambda \longrightarrow 0$ singularity is captured
by (\ref{lc_AA}) while the interference term is \beq \sum_{\rm
{\spins}}\left[{\cal M}{\bar{{\cal M}}}^\dagger+\bar{{\cal
M}}{{\cal M}}^\dagger\right] \,=\,a_0 \left[
2\frac{1+\alpha+\beta+2\lambda}{(\alpha+\lambda)(\beta+\lambda)}
-2\frac{1+\beta}{\left({\alpha+\lambda}\right){\beta}} \right],
\label{interf_lc} \eeq which is not singular. This shows that
light-cone gauge is indeed useful in identifying the $\alpha +
\lambda \longrightarrow 0$ singular terms.

It is also interesting to note the different source of the singularity
of the squared matrix element in the double logarithmic approximation. In the Feynman gauge the singularity
 originates in
the interference term: $1/\alpha$ appears because of the singularity
of the propagator of the quark with momentum $p+k$, and $1/\beta$
from the one with momentum $\bar{p}+k$. In the axial gauge the
singularity originates only from the diagram where the gluon
interacts with $p$, so $1/\alpha$ has the same source -- the
singularity of the propagator. On the other hand,~$1/\beta$ is
just built into the gauge~(\ref{propagator}).

\section{Dressing the gluon}

Let us demonstrate the application of the off-shell gluon splitting function for DGE. 
We begin with the
simplest example: the case of the massless\footnote{A perturbative analysis at NLL was
recently completed~\cite{CC} for the case of heavy quark fragmentation. Here we
discuss only the massless case.} quark fragmentation in $e^+e^-$ annihilation. The fragmentation
functions were analysed in the dispersive approach by Dasgupta and Webber~\cite{DasW_frag},
who concentrated on the determination of
the $x$ dependence of the power corrections. Our results for 
SDG characteristic function in the limit $x\longrightarrow 1$ are consistent with their calculation. 
The main difference between our approach and that of~\cite{DMW,DasW_frag}  
is that in the latter the power corrections 
are additive to the coefficient function, while here 
(similar ideas were raised before, see~\cite{Shape_function2}--\cite{Belitsky:2001ij}, 
\cite{BB_DY,AZ_KLN}, \cite{DW} and~\cite{DGE}), 
as we shall see in the next section, 
the SDG calculation is considered as the kernel of exponentiation, 
so the power corrections will appear as a function that is convoluted with the resummed 
coefficient function.   

The fragmentation functions are defined by
\beq
\frac{1}{\sigma}\frac{d\sigma}{dx\,d\cos\theta}=\frac38(1+\cos^2\theta)\,
F_T(x,Q^2)+\frac34\sin^2\theta \,F_L(x,Q^2)+\frac34\cos\theta \,F_A(x,Q^2),
\eeq
where $x\equiv 2qp_h/Q^2$. Here $q$ is the electroweak gauge boson momentum ($q^2\equiv Q^2>0$),
$p_h$ is the hadron momentum
and $\theta$ is the angle of the hadron with respect to the $e^-$ beam;
$F_T(x,Q^2)$, $F_L(x,Q^2)$ and $F_A(x,Q^2)$ correspond to the transverse, longitudinal and
asymmetric fragmentation functions,
respectively. $F_A(x,Q^2)$ vanishes for a pure electro-magnetic
interaction.
The standard factorization technique allows one to compute the coefficient function $C_P^i(z,Q^2)$ by
identifying the outgoing hadron as an outgoing parton~$i$. The physical fragmentation function is then
written as a convolution with the
non-perturbative function $D(x,Q^2)$,
\beq
F_P(x,Q^2)=\sum_i\int_x^1\frac{dz}{z} C_P^i(z,Q^2) D_i(x/z,Q^2),
\label{factor}
\eeq
where $i$ can be a quark, an antiquark or a gluon.
In moment space, \[\int_0^1 F_P(x,Q^2)x^{N-1}dx\equiv F_P(N,Q^2),\] the convolution translates into a product
\beq
F_P(N,Q^2)=\sum_i C_P^i(N,Q^2) D_i(N,Q^2).
\eeq

We will now show that the log-enhanced terms in the quark fragmentation coefficient function
$C^q(x,Q^2)$,
which are associated with a single gluon emission, can be directly calculated from~(\ref{splitting_function}).
In the next section this result will be used to calculate the multiple gluon (resummed)
coefficient function through DGE. 

The gluon virtuality $\epsilon$ has a crucial role in the calculation of the single gluon
emission cross-section to
all orders. For the calculation to be correct to any logarithmic accuracy (even in the simplest
case, i.e. the large-$N_f$ limit), an integral over the running coupling must be performed
with the correct argument -- the gluon virtuality. This is the physical scale of the interaction.
Using the dispersive approach in renormalon calculus, the integral over the running coupling is equivalent
to the sum of the perturbative series in the large-$\beta_0$ limit.
Such a renormalon calculation can be performed {\em within} the approximation considered in the derivation
of~(\ref{splitting_function}), where only the singular terms in the matrix element were kept.
This approximation amounts to keeping all the logs in the differential cross-section, while
neglecting terms that are explicitly suppressed by $1-x$.

The total quark fragmentation function $F(x,Q^2)=F_T(x,Q^2)+ F_L(x,Q^2)$ is
certainly sensitive to soft or collinear emission -- we saw that the squared 
matrix element (\ref{lc_AA}) (or
(\ref{full_me})) is singular for $\alpha+\lambda\longrightarrow 0$.
The quark that emits the gluon (in our gauge) is $p$. We therefore identify $p_h$ as $\bar{p}$, getting
\beq
x \equiv \frac{2\bar{p}q}{q^2}=\frac{1+\beta}{1+\alpha+\beta+\lambda}.
\eeq
We see that the singularity in (\ref{splitting_function}) is inversely proportional to $1-x\simeq
(\alpha+\lambda)/(1+\beta)$. Using this variable, the differential
gluon emission probability is
\beq
\frac{dP}{dx}\,=\,\frac{C_F\alpha_s}{2\pi} \,\int d\beta \left[ -\frac{\epsilon}{(1-x)^2}\,\frac{1}{(1+\beta)^2}
+\frac{1}{1-x}\,\frac{\beta^2+2\beta+2}{\beta(1+\beta)^3}\right],
\label{dP_dx}
\eeq
where $\epsilon=m^2/Q^2$.
In the integration over $\beta$, the limit that contributes to log-enhanced terms is the limit where the gluon
transverse momentum $\gamma$ vanishes. This translates into
\beq
\beta=\lambda/\alpha=\epsilon/(1-x-\epsilon).
\label{beta_limit}
\eeq
We therefore integrate over $\beta$ and substitute (\ref{beta_limit}), getting
\beq
\frac{dP}{dx}\,=\,\frac{C_F\alpha_s}{2\pi} \,{\cal F}(x,\epsilon)
\label{dP_dx1}
\eeq
with\footnote{Here we dropped sub-leading terms (e.g. ${\cal O}( {\epsilon^2}/{(1-x)^2})$),
which go beyond
our approximation and contribute only to non-logarithmic terms in the perturbative coefficients.}
\beq
\left. {\cal F}(x,\epsilon)\right\vert_{\log}\,=\,\frac{2}{1-x}\,\ln\left(\frac{1-x}{\epsilon}\right)\,-\,\frac{3}{2}\,\frac{1}{1-x}
\,+\,\frac{\epsilon}{(1-x)^2}\,+\,\frac12\,\frac{\epsilon^2}{(1-x)^3}.
\label{F_frag}
\eeq
The physical region is $\epsilon<1-x$, where~$\alpha$ and~$\beta$ are positive;
${\cal F}(x,\epsilon)$ is the off-shell gluon characteristic function~\cite{DMW,DasW_frag,BBM}
for the quark fragmentation function. The subscript `$\log$' in (\ref{F_frag})
is meant to remind us of the approximation: by
taking only the $1-x\longrightarrow 0$ singular terms in the matrix element,
we control only log-enhanced terms in the perturbative coefficients.
The full characteristic function (see e.g.~\cite{DasW_frag})
contains some additional terms, which generate non-logarithmic terms in the perturbative series.

In the dispersive approach~\cite{DMW} (see also \cite{BB,beneke,Average_thrust}) the single off-shell gluon
result (\ref{dP_dx1}) is promoted to
an infinite perturbative-sum associated with a single {\em dressed} gluon by replacing
the coupling with a dispersive integral
\beq
\frac{C_F}{2\beta_0}\,\int_0^{1-x}\frac{d\epsilon}{\epsilon}
 \dot{\cal F}(x,\epsilon)\,\bar{A}_{\eff}(\epsilon Q^2)\,=\,\frac{C_F}{2\beta_0}\,
\int_0^{1-x}\frac{d\epsilon}{\epsilon} \left[{\cal F}(\epsilon)-{\cal F}(0)\right]\,\bar{\rho}
(\epsilon Q^2),
\label{dispersive}
\eeq
where $\bar{A}(k^2)\equiv  \beta_0\bar{\alpha}_s(k^2)/\pi$.
The two integrals in (\ref{dispersive}) are related by integration by parts:
$\dot{\cal F}(x,\epsilon)\equiv-\epsilon{d{\cal F}(x,\epsilon)}/{d\epsilon}\,$,
and the function $\bar{\rho}(\mu^2)$ is
identified as the time-like discontinuity of the coupling,
\begin{equation}
\bar{\rho}(\mu^2) = \frac{1} {2\pi i}
 {\rm Disc}\left\{\bar{A}(-\mu^2)\right\}
\equiv
\frac{1}{2\pi i}\left[\bar{A}\left(-\mu^2+i\lambda\right)
-\bar{A}\left(-\mu^2-i\lambda\right)\right].
\label{discpt}
\end{equation}
The ``time-like coupling'' $\bar{A}_{\eff}(\mu^2)$ in (\ref{dispersive}) obeys
$\mu^2\,{d\bar{A}_{\eff}(\mu^2)}/{d\mu^2}=\bar{\rho}(\mu^2)$.
For example, in the one-loop case $\bar{A}(k^2)=1/(\ln k^2/\bar{\Lambda}^2)$ and
\[
\bar{A}_{\eff}(\mu^2)
=\frac12-\frac{1}{\pi}\arctan\left(\frac{1}{\pi}\ln\frac{\mu^2}{\bar{\Lambda}^2}\right).
\label{1_loop_time_like}
\]
We emphasize that (\ref{dispersive}) is considered just as a perturbative sum, and {\em not} as a
non-perturbative definition\footnote{Such a definition differs from the Borel-sum
by power-corrections which are not associated with the infrared 
dynamics~\cite{Average_thrust,BB,beneke}.}. 

The bar over the coupling indicates 
a specific renormalization scheme. In the large-$\beta_0$ limit (large-$N_f$ limit)
$\bar{A}(k^2)$ is related to the $\rm {\overline{MS}}$ coupling
{${A_{\MSbar}}(k^2)\equiv \beta_0{\alpha}^{\MSbar}_s(k^2)/\pi$} by
$\bar{A}(k^2)={A_{\MSbar}}(k^2)/\left({1-\frac53 {A_{\MSbar}}(k^2)}\right)$.
In the general case, going beyond this limit requires a skeleton
expansion~\cite{Disentangling}, generalizing the Abelian one.
However, in the context of our approximation, where only the singularity of the splitting 
function matters,  a skeleton expansion is not required.
The appropriate coupling is the ``gluon bremsstrahlung'' effective charge~\cite{CMW,DGE}.
Fixing $\bar{A}$ by
\beq
\bar{A}(k^2)=\frac{{A_{\MSbar}}(k^2)}{1-\left[\frac53
+\left(\frac13-\frac{\pi^2}{12}\right)C_A/\beta_0\right] {A_{\MSbar}}(k^2)}
\label{A_bar_brem}
\eeq
the splitting function~(\ref{splitting_function}) is correct to next-to-leading order, as far as the
singular terms are concerned. The origin of the ``gluon bremsstrahlung'' effective charge can be understood
by the formulation of the Sudakov resummation through an evolution equation for
Wilson-line operators. 
In this language this effective charge appears as the cusp anomalous dimension~\cite{Korchemsky:1989si,Korchemsky:1993xv}.

In the quark fragmentation case~(\ref{F_frag}), we have
\beq
\left. {\dot {\cal F}}(x,\epsilon)\right\vert_{\log}\,
=\,\frac{2}{1-x}\,-\,\frac{\epsilon}{(1-x)^2}\,
-\,\frac{\epsilon^2}{(1-x)^3}.
\label{dF_frag}
\eeq
Note that ${\dot {\cal F}}(x,\epsilon=0)\neq 0$ so the integral (\ref{dispersive}) diverges for
vanishing gluon virtuality. This is a direct consequence of the fact
that the fragmentation function is
not collinear-safe. Factorization~(\ref{factor}) is
required. We shall make the following identification:
$C^q(x,Q^2)=\delta(1-x)+C_{\SDG}^q(x,Q^2)+\ldots$ where the ellipses stand for corrections that are sub-leading by
powers of either $\beta_0$ or $1-x$ and 
\begin{eqnarray}
\label{Cq}
 &&\!\!\!\!\!\!\!\!\!\!\!\! C_{\SDG}^q(x,Q^2)=\frac{C_F}{2\beta_0}\,\left[
\,\int_1^{1-x}\frac{d\epsilon}{\epsilon}
 {\cal S}(x)\,\bar{A}_{\eff}(\epsilon Q^2)\,+\,\int_0^{1-x}\frac{d\epsilon}{\epsilon}
 \left(\dot{\cal F}(x,\epsilon)-{\cal S}(x)\right)\,\bar{A}_{\eff}(\epsilon Q^2)\,\right]
 \\ \nonumber
&&=\frac{C_F}{2\beta_0}\,\left[\frac{2}{1-x}\,\int_{1}^{1-x} \,\frac{d\epsilon}{\epsilon}
 \,\bar{A}_{\eff}(\epsilon Q^2)\,
-\,\int_{0}^{1-x} \,\frac{d\epsilon}{\epsilon}
\,\left(\frac{\epsilon}{(1-x)^2} \,
+\,\frac{\epsilon^2}{(1-x)^3}\right)\,\bar{A}_{\eff}(\epsilon Q^2)
\right].
\end{eqnarray}
Here ${\cal S}(x)=2/(1-x)$ is the coefficient of $\ln ({1}/{\epsilon})$ in ${\cal F}(x,\epsilon)$.
The subtracted term is absorbed into the non-perturbative
function $D_q(x/z,Q^2)$.
This means that its evolution is controlled by ${\cal S}(x)$, i.e. by $C_F(\bar{\alpha}_s/\pi)/(1-x)$.
With the definition~(\ref{Cq}) the coefficient function has a well-defined
perturbative expansion. Expanding the coupling $\bar{A}_{\eff}(\epsilon Q^2)$ in
a fixed-scale coupling
one can write the corresponding perturbative coefficients in terms of the log-moments:
\begin{eqnarray}
h_n(x) \,&\equiv&\,
\,\int_{1}^{1-x} \,\frac{d\epsilon}{\epsilon}
 \left(\ln\frac{1}{\epsilon}\right)^n {\cal S}(x)+\int_{0}^{1-x} \,\frac{d\epsilon}{\epsilon}
 \left(\ln\frac{1}{\epsilon}\right)^n \left(\dot{\cal F}(x,\epsilon)-{\cal S}(x)\right)\\
 \nonumber
 &=&\frac{-1}{(1-x)}
\left[ \frac{2}{n+1} L^{n+1}
\,+\, e^L\Gamma\left(n+1,L\right)\,+\, 2^{-n-1}\, e^{2L}\Gamma\left(n+1,2L\right)
\right],
\label{log_moments}
\end{eqnarray}
where $L=\ln 1/(1-x)$. Since $e^L\Gamma(n+1,L)=\sum_{j=0}^n L^j n!/j!$, the expression in the 
square brackets is just a polynomial of order $n+1$ in $L$. Note, however, that since the
sub-leading logs increase factorially, 
the sum is not well approximated by the leading log term even at rather large $L$
(see~\cite{DGE}).

Contrary to the total fragmentation function, the longitudinal
fragmentation function $F_L(x,Q^2)$
is insensitive to soft or collinear emission (at the logarithmic level):
projecting (\ref{Tr_ee}) onto $\epsilon_L^\nu \epsilon_L^\rho$, where
\[
\epsilon_L^\nu\equiv\frac{1}{\vert{\bar{\bf p}}\,\vert}
\left(\bar{p}^\nu\,-\,\frac{q\bar{p}}{q^2}\,q^\nu\right)
\]
is the polarization along the direction of $\bar{\bf p}$, the contribution vanishes. 
It follows that in the electromagnetic case the log-enhanced terms in the
total fragmentation function all originate in the transverse fragmentation function 
(see~\cite{DasW_frag,BBM}). Exactly the same log-enhanced terms appear in the parity-violating
asymmetric fragmentation function.

\section{Dressed gluon exponentiation}

We saw that factorization of the squared matrix element, which is the basis for exponentiation,
holds also for an off-shell gluon.
Exponentiation can be derived, as in \cite{Contopanagos:1997nh,CTTW}, from an evolution equation, or, as
in~\cite{DGE}, directly from the probabilistic interpretation of the SDG cross-section based on 
the approximation of independent emission.

The factorization of the squared matrix element suggests that
the entire log-enhanced SDG sum exponentiates. In
reality straightforward exponentiation holds only up to a 
certain logarithmic accuracy, because
gluons are correlated through the phase-space
integration. Momentum conservation translate into a 
factorized form in Laplace space~(see e.g.~\cite{CTTW,DGE}). 
To some logarithmic accuracy one can then perform the
integration over the gluon momenta as if other 
gluons had not been emitted. 
At the same time exponentiation of the entire SDG
sum can be regarded as the leading contribution to the 
{\em exponent} in the large-$\beta_0$ limit. In this sense the expansion in powers
of $1/\beta_0$ in the exponent replaces the standard expansion in powers of the log.

Being aware of this we now proceed to write down the
explicit exponentiation formula for the fragmentation function
case. We work in the approximation where the invariant mass of the jet, $Q^2(1-x)$, is additive with
respect to multiple gluon emission. With this assumption we ignore
correlations between the gluons, which would modify the
exponentiation beyond NLL accuracy. The resummed expression
for~$C^q(x,Q^2)$ is then given by the inverse Mellin transform of
$\left.C_q(N,Q^2)\right\vert_{\res}$, where \beq
\left.C_q(N,Q^2)\right\vert_{\res}=\exp\,\left\{\int_0^1\, dx\,
\left(x^{N-1}-1\right)\,C_{\SDG}^q(x,Q^2)\right\}.
\label{expo}
\eeq
Here we drop an $x$-independent factor associated with the virtual corrections 
to the hard cross-section~(see e.g.~\cite{CC}).

A convenient way to deal with perturbation theory to all orders is
Borel summation. Starting with the scheme-invariant Borel
representation of the coupling~\cite{BY}--\cite{G3}, \beq
\bar{A}(k^2)=\int_0^{\infty}d{u} \,\exp\left({-{u} \ln
k^2/\Lambda^2}\right) \bar{A}_B({u}) \label{A_Borel} \eeq and
taking the time-like discontinuity \cite{BY} we get \beq
\bar{A}_{\eff} ( Q^2)=\int_0^{\infty}d{u}\,\exp\left(-{{u} \ln
Q^2/\Lambda^2}\right)\,
 \frac{\sin\pi{u}}{\pi{u}}\, \bar{A}_B({u}).
\label{Borel_A_eff}
\eeq
Using this Borel representation of $\bar{A}_{\eff} (\epsilon Q^2)$, and integrating (\ref{Cq}) over
$\epsilon$, one obtains a Borel integral representing the SDG perturbative sum:
\beq
C_{\SDG}^q(x,Q^2)
=\frac{C_F}{2\beta_0} \,\int_0^{\infty} d{u} B_{\SDG}({u},x) \,\exp\left(-{{u} \ln Q^2/\Lambda^2}\right)\,
 \frac{\sin\pi{u}}{\pi{u}}\, \bar{A}_B({u})
\label{Borel_rep}
\eeq
with the following Borel function
\beq
B_{\SDG}(u,x) = \frac{-1}{1-x}\,\left[\frac{2\left((1-x)^{-u}-1\right)}{u}
+\frac{(1-x)^{-u}}{1-{u}}+\frac{(1-x)^{-u}}{2-{u}}
 \right].
\label{Borel}
\eeq
Next, the exponent in Mellin space
$\ln \left.C_q(N,Q^2)\right\vert_{\res}$ is obtained from (\ref{expo}) integrating over~$x$
\beq
\ln \left.C_q(N,Q^2)\right\vert_{\res}
=\frac{C_F}{2\beta_0} \,\int_0^{\infty} d{u} B_{N}({u}) \,\exp\left(-{{u} \ln Q^2/\Lambda^2}\right)\,
 \frac{\sin\pi{u}}{\pi{u}}\, \bar{A}_B({u})
\label{Borel_rep_N}
\eeq
with\footnote{Upon integration $\Gamma(-u)$ appears with a factor $\Gamma(N)/\Gamma(N-u)$. In the approximation
considered ($N\gg 1$) this factor can be replaced by $e^{u\ln N}$.}
\beq
B_N({u}) \,=\frac{-2}{u}\left[e^{u\ln N}\Gamma(-u)+\frac1u+\gamma_E+\ln
N\right]-\left(\frac{1}{1-u}+\frac{1}{2-u}\right)\left[e^{u\ln N}\Gamma(-u)+\frac1u \right].
\label{exp_Bor}
\eeq

The Borel function~(\ref{exp_Bor}) contains valuable information on both the
perturbative and the non-perturbative levels.

{\bf Sudakov logs:}
To get an explicit perturbative expansion of the exponent in Mellin space one can perform
the Borel integral term by
term, similarly to what was done in~\cite{DGE} for the case of the thrust distribution.
For simplicity we work with a one-loop running coupling~${\bar{A}(Q^2)}$.
It is straightforward to rewrite the sum as
\beq
\ln \left.C_q(N,Q^2)\right\vert_{\res}=\frac{C_F}{2\beta_0}
 \sum_{k=1}^{\infty}{{\bar{A}(Q^2)}^{k-2}}\,\, f_{k}\!\left({\bar{A}}(Q^2)\ln N\right),
\label{log_expansion}
\eeq
where the functions $f_k(\xi)$ sum all powers of $\xi\equiv {\bar{A}}(Q^2)\ln N$.
They are given by
\beq
f_k(\xi)=\sum_{l=0}^{(k-1)/2}\,\frac{(-\pi^2)^l}{(2l+1)!}\,g_k^{(l)}(\xi)\,M_{\rm col}(k),
\label{f_k}
\eeq 
where $g_k^{(l)}(\xi)$ is the following hypergeometric function
\begin{eqnarray}
g_k^{(l)}(\xi)=\xi^{n_0}\,\,\frac{\Gamma(k-2+n_0)}{\Gamma(1+n_0)}\, 
_2F_1\left( [1, k-2+n_0],[1+n_0],\xi\right)
\label{g_kl}
\end{eqnarray}
with $n_0=\max(0,2l+3-k)$ and
\begin{eqnarray}
\label{M_col}
M_{\rm col}(k)=2\,c_{k-2}+\sum_{m=0}^{k-2}(1+2^{-m-1})\,c_{k-3-m},
\end{eqnarray}
where the numbers $c_k$ are defined by $\Gamma(-{z})=-\sum_{k=-1}^{\infty}\,c_k\,{z}^k$. This means that
$c_{-1}=1$, $c_0=\gamma_E$, $c_1={\pi^2}/{12}\,
+\,{\gamma_E^2}/{2}$, etc. Higher $c_k$ contain higher $\zeta_i$ numbers ($i\leq k+1$),
yet numerically $c_k$ ($k\geq 1$) are all close to $1$.
For $k=1,2$ the functions are
\begin{eqnarray}
\label{f_12}
f_{1}(\xi)&=& 2\,\left(\xi+(1-\xi)\ln(1-\xi)\right)\\ \nonumber
f_2(\xi) &=& -\left(\frac32+2\gamma_E\right)\ln(1-\xi)
\end{eqnarray}
while for $k\geq 3$ they are characterized by increasing coefficients as well as increasing order
pole singularities at $\xi=1$,
\begin{eqnarray}
\label{f_k_expml1}
\begin{array}{ll}
f_{3}(\xi)=   { 0.804}/{(1-\xi ) } + 3.29 \,\xi   \\
f_{4}(\xi)= {0.779}/{({1-\xi  } )^{2}}   \\
f_{5}(\xi)=   { 2.32}/{({1-\xi  })^{3}}  -9.74\,\xi \\
f_{6}(\xi)= {6.12}/{({1-\xi  } )^{4}}   \\
f_{7}(\xi)=  { 23.8}/{({1-\xi  } )^{5}} +45.78\,\xi  \\
f_{8}(\xi)=  {120.9}/{({1-\xi  } )^{6}}  \\
f_{9}(\xi)=721.54/(1-\xi)^7-263.57\,\xi \\
f_{10}(\xi)=5043.48/(1-\xi)^8.
\end{array}
\end{eqnarray}
We see that sub-leading logs are characterized by an explicit
factorial growth as well as by an increasing singularity at the
end of the perturbative region: \beq \xi\equiv
\frac{\alpha_s\beta_0}{\pi}\,\ln \frac{1}{1-x}\,\,=\,\,\xi_{\max},
\eeq where $\xi_{\max}=1$. A similar structure was identified in
the case of the thrust (or jet mass) distribution in~\cite{DGE}.
The two cases are compared in the next section. As discussed
in~\cite{DGE} the enhancement of sub-leading logs implies that an
approximation based on a fixed logarithmic accuracy in the
exponent has a small range of validity, namely $\xi\ll
\xi_{\max}$. The resummed exponent (\ref{Borel_rep_N}) is valid
much closer to the strict limit of applicability of perturbation
theory ($\xi\lsim \xi_{\max}$), provided appropriate
power corrections are included.

{\bf Power corrections:} The form of the power corrections can be
deduced from the renormalon ambiguity. They are
expected to be additive at the level of (\ref{Borel_rep_N}) and
therefore they exponentiate together with the perturbative sum, so that 
\beq
C_q(N,Q^2)=\left.C_q(N,Q^2)\right\vert_{\res}\,C_q^{\rm NP}(N,Q^2).
\eeq 
This implies that the correction can be written as a convolution in $x$ space.
The Borel singularity of the fragmentation function exponent 
in the large-$\beta_0$ limit can be read off eq.~(\ref{exp_Bor}), taking into
account the attenuation by the $\sin(\pi u)$ factor in
(\ref{Borel_rep_N}). The result turns out to be very simple: a
pole at $z=1$ with a residue $-N$ and second pole at $z=2$ with a
residue $-N^2/4$. The non-perturbative correction $C_q^{\rm NP}(N,Q^2)$ 
is therefore expected to~be
\beq
C_q^{\rm NP}(N,Q^2) = 
\exp\left\{-\omega_1\frac{C_F}{2\beta_0}\frac{N\bar{\Lambda}^2}{Q^2}
-\omega_2\frac{C_F}{2\beta_0}\frac{N^2\bar{\Lambda}^4}{4Q^4}\right\},
\label{C_NP}
\eeq
where $\omega_n$ are dimensionless constants. Since $C_q(N,Q^2)$ multiplies a
non-perturbative distribution $D_q(N,Q^2)$, eq.~(\ref{C_NP}) leads to a
meaningful prediction only if it dominates the behaviour in the threshold
region. This possibility should be tested experimentally. 
As discussed in the next section, the situation is different for
collinear- and infrared-safe quantities where no non-perturbative 
distribution is needed a priori.

\section{Jet mass in $e^+e^-$ annihilation}

One of the classical examples where soft and collinear gluon
radiation is important is the case of event-shape variables in
$e^+e^-$ annihilation. As opposed to the fragmentation function
case, these observables are infrared- and collinear-safe, so the
perturbative calculation does not require factorization. The
enhanced sensitivity of these observables to large angle soft
emission is reflected both perturbatively -- as large
logs~\cite{CTTW}, and non-perturbatively -- as
power corrections~\cite{Average_thrust,DGE} (see also 
\cite{DMW}, \cite{wise}--\cite{higher_moments} and \cite{Shape_function2}--\cite{Belitsky:2001ij}).

Here we consider event-shape variables that are related to the jet mass distribution, such as the thrust and the
heavy jet mass. The thrust distribution was analysed in detail, using DGE, in~\cite{DGE}. Analysis of the
heavy jet mass by the same method will appear soon~\cite{Heavy_Jet}.
The starting point in~\cite{DGE} was the {\em full} SDG characteristic function~\cite{higher_moments}.
We shall now see how the relevant part of the characteristic function,
the one that generates the logs, can be computed directly from the generalized off-shell
splitting function. This will be followed by a brief description of the results of~\cite{DGE} for
comparison with other cases studied here.

Consider a single off-shell gluon emission, which is either soft or collinear (two-jet kinematics).
Assuming that the gluon is in the hemisphere of the quark $p$, it is $\bar{p}$ that sets the thrust axis.
The $\bar{p}$ jet mass is zero and the $p$ jet mass, which is also $1\,-$ thrust, is simply
\beq
\rho=\frac{m_H^2}{Q^2}=\frac{(p+k)^2}{Q^2}=\frac{2pk+k^2}{Q^2}=\frac{(\alpha+\lambda)\,2p\bar{p}}{Q^2}
=\frac{\alpha+\lambda}{1+\alpha+\beta+\lambda}=1-x.
\label{rho_x}
\eeq
Thus calculation of the SDG characteristic
function (neglecting non-inclusive effects~\cite{DMW,Average_thrust,higher_moments,DGE},
associated with gluon decay into opposite hemispheres) is just {\em identical} to that of the
total fragmentation function. The result is (\ref{dF_frag}), with the replacement of $1-x$ by $\rho$:
\beq
\left. \dot{\cal F}(\rho,\epsilon)\right\vert_{\log}=\frac{2}{\rho}-\frac{\epsilon}{\rho^2}-\frac{\epsilon^2}{\rho^3}.
\label{dF_jets}
\eeq
This result is confirmed by the explicit calculation of the full characteristic 
function~\cite{higher_moments} followed by the identification of terms that contribute 
to logs~\cite{DGE}.

A major difference between jet mass distribution and the fragmentation function case is
in the large-angle phase-space limit:
we assumed that the gluon is in the hemisphere of the quark $p$. This is equivalent to
$\beta>\alpha$. Along the zero transverse momentum boundary of phase space~$\alpha\beta=\lambda$,
this condition translates into $\beta>\sqrt{\lambda}$.
It follows that the lower integration limit over $\epsilon$ is $\epsilon=\rho^2$, which guarantees 
infrared-safety. The upper integration limit is $\epsilon=\rho$, as for the fragmentation 
function case. The SDG differential cross-section is therefore
\beq
\left.\frac{1}{\sigma}\frac{d\sigma}{d\rho}(Q^2,\rho)\right\vert_{\SDG}=
\frac{C_F}{2\beta_0}\int_{\rho^2}^{\rho}\frac{d\epsilon}{\epsilon}\,\bar{A}_{\eff}
(\epsilon Q^2) \,\dot{\cal F}(\rho,\epsilon),
\label{ours}
\eeq
where $\dot{\cal F}(\rho,\epsilon)$ is given by (\ref{dF_jets}).

Proceeding along the lines of the previous section we first use~(\ref{Borel_A_eff}) to
write a Borel representation of the SDG cross-section. The Borel function
is~\cite{DGE}
\beq
B_{\SDG}(u,\rho)= \frac1\rho\left[\frac{2}{u} \exp\left(2u\ln\frac 1\rho\right)
-\left(\frac{2}{u}+\frac{1}{1-u}+\frac{1}{2-u}\right)
 \exp\left({u} \ln\frac 1\rho\right)\right].
\eeq
Next, the exponent in Laplace space~\cite{CTTW,DGE} is
\beq
\ln J_{\nu}(Q^2)=\int_0^1 \left.
\frac{1}{\sigma}\frac{d\sigma}{d\rho}(Q^2,\rho)\right\vert_{\SDG}
\left(e^{-\nu \rho}-1\right)d\rho.
\eeq
The corresponding Borel function is given by
\begin{eqnarray}
\label{Borel_nu}
B_{\nu}(u)&=& \int_0^1 \frac{d\rho}{\rho} \left[\frac{2}{u}
e^{2u \ln\frac1\rho}-\left(\frac{2}{u}+\frac{1}{1-u}+\frac{1}{2-u}\right)
 e^{u \ln\frac1\rho} \right]\,\left(e^{-\nu \rho}-1\right)\\
&\simeq&\frac{2}{u}\left[e^{2u\ln\nu}\Gamma(-2u)
+\frac{1}{2u}\right]-\left(\frac{2}{u}+\frac{1}{1-u}
+\frac{1}{2-u}\right)\left[e^{u\ln \nu}\Gamma(-u)
+\frac{1}{u}\right],\nonumber
\end{eqnarray}
where non-logarithmic terms were neglected. It is interesting to
compare this result with the fragmentation function
case~(\ref{Borel_rep_N}) and~(\ref{exp_Bor}), where only the
collinear limit $\epsilon=1-x=\rho$ plays a role. The terms asociated with
 the collinear limit are common. 
Equation~(\ref{Borel_nu}) contains in addition a term that corresponds
to large-angle emission~$\epsilon=\rho^2$. The latter has a double
argument $u\longrightarrow 2u$, indicating larger perturbative
coefficients as well as enhanced power-corrections~\cite{DGE}.

Equation~(\ref{Borel_nu}) summarizes the DGE result for the 
jet-mass distribution. Performing the Borel integral as
in~(\ref{Borel_rep_N}) and the inverse Laplace transform, this
result can be directly used for phenomenological analysis of the
thrust and heavy jet mass distribution~\cite{DGE,Heavy_Jet}. A
detailed analysis of the perturbative structure of the exponent
(sub-leading logs) and of the expected power corrections was
performed in~\cite{DGE}. The main results are the following.

{\bf Sudakov logs:} Performing the Borel integral term by term with the one-loop running coupling we
can write the exponent $\ln J_{\nu}(Q^2)$ as in eq.~(\ref{log_expansion}) with
\beq
f_k(\xi)=\sum_{l=0}^{(k-1)/2}\,\frac{(-\pi^2)^l}{(2l+1)!}
\,\left[g_k^{(l)}(\xi)\,M_{\rm col}(k)\,+\,g_k^{(u)}(2\xi)\,M_{\rm la}(k)
\right]
\label{f_k_Jet}
\eeq
where $\xi\equiv \bar{A}(Q^2)\ln\nu$ and $g_k^{(l)}(\xi)$ is given by (\ref{g_kl}).
$M_{\rm col}(k)$ (given by (\ref{M_col})) and 
\begin{eqnarray}
\label{M_ls}
M_{\rm la}(k)&=&-2^{k-1}\,c_{k-2}
\end{eqnarray}
correspond to the collinear and large-angle contributions to the Borel function, respectively.
Explicitly, this yields
\begin{eqnarray}
\label{f_k_expml}
f_{1}(\xi)&=& 2 (1- \,\xi)\,\ln(1 - \xi) - (1 - 2\,\xi)\,
\ln(1 - 2\,\xi)   \nonumber \\
f_{2}(\xi)&=& - 2\,\gamma_E \,(\ln(1 - \xi) - \ln(1 - 2\,\xi)) -
{\  \frac {3}{2}} \,\ln(1 - \xi)
\end{eqnarray}
and
\begin{eqnarray*}
\begin{array}{ll}
f_{3}(\xi)=   { 0.804}/{(1-\xi ) }  - {  {2.32}/({1 -2\xi   })}  \\
f_{4}(\xi)= {0.779}/{({1-\xi  } )^{2}} - {  {2.68}/{(1 -2\xi )^{2}}}  \\
f_{5}(\xi)=   { 2.32}/{({1-\xi  })^{3}}  - {  {5.00}/{(1 -2\xi )^{3}}} \\
f_{6}(\xi)= {6.12}/{({1-\xi  } )^{4}}  - { {15.32}/{(1 -2\xi )^{4}}} \\
f_{7}(\xi)=  { 23.8}/{({1-\xi  } )^{5}}  - {  {61.12}/{(1 -2\xi )^{5}}} \\
f_{8}(\xi)=  {120.9}/{({1-\xi  } )^{6}} - {  {305.52}/{(1 -2\xi )^{6}}}, \\
f_{9}(\xi)=  {721.54}/{({1-\xi  } )^{7}}  - { 1833.55/{(1 -2\xi )^{7}}} \\
f_{10}(\xi)=  {5043.48}/{({1-\xi  } )^{8}} - {  12834/{(1 -2\xi )^{8}}},
\end{array}
\end{eqnarray*}
The factorially increasing coefficients of $f_k(\xi)$ and the
enhanced singularity at $\xi=1/2$ imply that a fixed
logarithmic accuracy calculation holds only in a very restricted
region $\xi\ll 1/2$. Otherwise, these sub-leading logs must be
resummed~\cite{DGE}. Having performed such resummation and having
included the appropriate power corrections (see below) the region
of applicability of the result can be extended up to
$\xi<\xi_{\max}=1/2$. We saw that in the absence of large angle
soft emission sensitivity (the fragmentation function case) this
region is larger: $\xi<\xi_{\max}=1$.

{\bf Power corrections:} The ambiguity of the Borel integral indicates specific power
corrections~\cite{DGE}. Taking into account the singularities of (\ref{Borel_nu}) and the factor
$\sin \pi u/\pi u$ of eq.~(\ref{Borel_A_eff}), the Borel integrand (in the
large-$\beta_0$ limit) is singular at half
integers~\hbox{$u=\frac12,\frac32, \frac52 \ldots$,}
 and at $u=1,2$. The first type of singularity is a result of
the large angle soft emission (the first term in (\ref{Borel_nu})). It
translates to power corrections of the form $\sim (\nu/Q)^n$ where $n$ is an {\em odd} integer,
\beq
J_\nu^{\rm NP}(Q^2)=\exp\left\{\sum_{i=1}^{\infty}\frac{2C_F\,(-1)^{i+1}}{\pi{\beta_0}\,(2i-1)^2\, 
(2i-1)!}\,\omega_{(2i-1)/2} \,\left(\frac{{\bar{\Lambda}}\nu}{Q}\right)^{2i-1}\right\}.
\label{J_nu_NP}
\eeq
The large angle power corrections are important when~$\rho\sim1/\nu$ approaches~$\bar{\Lambda}/Q$.
They can be resummed into a non-perturbative shape function~\cite{Shape_function1}--\cite{DGE}, corresponding
to the inverse Laplace transform of~(\ref{J_nu_NP}).
Thus, the calculation of the exponent in
the large-$\beta_0$ limit by DGE yields highly non-trivial information on the shape function: its even
central moments (controlling corrections $\sim (\nu/Q)^n$ with even~$n$) are suppressed~\cite{DGE}.
The second type of singularity is a result of collinear emission. As in the fragmentation function case, 
it translates into power corrections of the
form $\sim 1/(Q^2\rho)$ and  $\sim 1/(Q^4\rho^2)$. These corrections can be neglected in the 
region $\rho Q/\gsim \bar{\Lambda}$.

\section{Deep inelastic structure functions}

Deep inelastic structure functions differ from the previous
examples by the fact that they have a systematic expansion in the
powers of the hard scale -- the twist expansion~\cite{EFP,JS}. In
the region of interest, $x_{\rm Bj}=Q^2/(2Pq)\longrightarrow 1$
(here $P$ is the nucleon momentum and $q$ is the virtual photon
momentum ($-q^2\equiv Q^2$))  the twist expansion breaks down and
power corrections of the form $1/Q^{2n}(1-x_{\rm Bj})^n$ become
dominant. In this respect, the situation is analogous to the
fragmentation function case.

The renormalon approach was used to study the $x_{\rm Bj}$ dependence of
power corrections~(see \cite{DasW_DIS}--\cite{Akhoury:1997rt} and \cite{DMW}).
The correspondence between the renormalon approach and the twist expansion was addressed in~\cite{beneke}
in the specific case of the longitudinal structure function $F_L$. It was shown there that the infrared
renormalon ambiguity of the twist-two coefficient functions cancels against the ambiguity of the twist-four
matrix elements, ambiguity that results from the ultraviolet quadratic divergence of these matrix elements.
The case of the transverse structure function $F_2$ is more interesting 
than that of $F_L$, owing to the threshold region~\cite{Korchemsky:1993xv}.
The matching of the renormalon ambiguity and the twist expansion in this case is discussed in~\cite{DIS}.

Contrary to the $e^+e^-$ examples analysed above, in the case of deep inelastic scattering,
both incoming and outgoing quarks can be the source of gluon bremsstrahlung.
Nevertheless, in the approximation considered (and in the appropriate gauge) one can
describe the radiation as if it were associated with the {\em final-state quark} alone. Indeed, we will
see that the SDG characteristic function, which generates the log-enhanced part of the $F_2$
coefficient function, is identical to that of the quark fragmentation function.

Identifying the incoming quark momentum $P$ with $\bar{p}$ (``$-$'' direction) and fixing the
axial gauge $A_+=0$ we consider gluon emission ($k$) off the outgoing quark, which carries
momentum $p$ in the ``+'' direction. In this gauge the gluon coupling to the incoming quark does
not give rise to singular terms. Momentum conservation implies that
$\bar{p}+q=p+k$. The standard $x_{\rm Bj}$, expressed in terms of the light-cone
parameters~(\ref{lc_par}) is
\beq
x_{\rm Bj}=Q^2/(2Pq)=Q^2/(2\bar{p}q)=1-(\alpha+\lambda)/(1+\beta).
\eeq

The matrix element for emission off the outgoing quark $p$ is~(\ref{M22}).
The phase-space integral is
\begin{eqnarray}
\label{PS_DIS}
&&\int \frac{d^4k}{(2\pi)^3}\,\delta\left(k^2-m^2\right)
d^4p\,\delta\left(p^2\right)\,\delta^4(\bar{p}+q-k-p)  \\ \nonumber
&=&\frac{1}{8\pi^2}\int\,
\frac{d\alpha\,d\beta}{(1+\beta)^2}\,
\delta\left(\alpha-(1-x_{\rm Bj}-\epsilon x_{\rm Bj})(1+\beta)\right),
\end{eqnarray}
where $\epsilon\equiv k^2/Q^2=\lambda/(x_{\rm Bj}(1+\beta))$.
Expressing~(\ref{M22}) in terms of $x_{\rm Bj}$ and $\beta$, we
find that the partonic differential cross-section for a single
gluon emission coincides with (\ref{dP_dx}) 
upon replacing $1-x$ by $1-x_{\rm Bj}$. As in the $e^+e^-$
case, the integration over $\beta$ is bounded from below by the
positivity of the squared transverse momentum. For vanishing
transverse momentum $\alpha\beta=\lambda$ one finds \beq
\beta=\frac{\epsilon x_{\rm Bj}}{{1-x_{\rm Bj}-\epsilon x_{\rm
Bj}}}\simeq \frac{\epsilon}{1-x_{\rm Bj}-\epsilon}.
\label{DIS_beta_limit} \eeq Integrating the gluon emission
probability over $\beta$ and substituting the limit
(\ref{DIS_beta_limit}), one gets the following characteristic
function \beq \left. {\dot {\cal
F}}(x,\epsilon)\right\vert_{\log}\,
=\,\frac{2}{1-x_{\rm Bj}}\,-\,\frac{\epsilon}{(1-x_{\rm Bj})^2}\,
-\,\frac{\epsilon^2}{(1-x_{\rm Bj})^3}, \label{dF_frag_DIS} \eeq which
is {\em identical} to the fragmentation function
case~(\ref{dF_frag_DIS}). This
result can be compared with the full characteristic function
computed in~\cite{DasW_DIS,DMW}. The additional terms in the
latter do not contribute to log-enhanced terms in the perturbative
coefficients. In the approximation considered, the integration
over the gluon virtuality has the same range as in the
fragmentation function case, i.e.~$\epsilon<1-x_{\rm Bj}$. Also
here there is no restriction from below (the coefficient
functions are not collinear-safe) so that factorization into a
non-perturbative parton distribution must be used. It follows that
the structure of the exponent in the large-$\beta_0$ limit is just
identical to that of the fragmentation function case, as
summarized by eqs.~(\ref{Borel_rep_N}) and~(\ref{exp_Bor}). As
discussed at the end of section~6, this structure has definite
implications in both the perturbative\footnote{The resummation of
deep inelastic coefficient function was recently
performed~\cite{Vogt} to NNLL accuracy. Our results are consistent
with those calculations.} and non-perturbative levels. The fact that the exponent 
is identical to the fragmentation function case suggests that it 
is the general properties of the quark jet evolution 
that control the power corrections and not the particular process 
from which the jet emerges. 

\section{Initial-state radiation in the Drell--Yan process}

The classical example where initial-state radiation determines the cross-section near threshold
 is the case of the
Drell--Yan process, the inclusive cross-section for lepton pair
production in hard hadronic collisions. Resummation of the leading
and next-to-leading logs for this process in the limit where
$x\equiv Q^2/(p+\bar{p})^2$ approaches $1$ was
performed~\cite{CT_DY,Contopanagos:1997nh,CMW} and the associated
power corrections were discussed
in~\cite{CS_DY,Shape_function1,BB_DY,AZ_KLN}.

The factorization property of soft and collinear gluons holds for
initial-state radiation exactly as for final-state radiation. In
particular, at the level of the squared matrix element, the
emission of soft or collinear gluons appears as an overall factor,
multiplying the cross-section for no emission, as in
eq.~(\ref{M22}). However, as we shall see below, the way the
singularity of the propagator is related to the physical parameter
that controls the phase space, $1-x$, is fundamentally different
and consequently so is the structure of the exponent. Although the
leading and next-to-leading logs are similar to those of the jet
mass (or the thrust) distribution in $e^+e^-$ annihilation, the
characteristic function and thus the sub-leading logs and the
power corrections~\cite{BB_DY} are different. 

Consider now the case of quark ($p$) antiquark ($\bar{p}$)
annihilation into an off-shell boson~$q$ (a photon in Drell--Yan),
where the quarks originate in the colliding hadrons. In the
perturbative calculation, we begin with on-shell partons
$p^2=\bar{p}^2=0$. Choosing the frame such that $p$ is in the
``$+$'' direction and the gauge $A_{+}=0$, so that the radiation
decouples from $\bar{p}$, we start by calculating the squared
matrix element for a single gluon emission off the quark $p$.
The gluon momentum $k$ is decomposed as in~(\ref{lc_par}).
Using~(\ref{M2_DY}) and the following phase-space integral
\begin{eqnarray*}
\int \frac{d^4k}{(2\pi)^3}\,\delta\left(k^2-m^2\right)
d^4q\,\delta\left(q^2-2p\bar{p} \,x\right)\,\delta^4(q+k-p-\bar{p})
=\int\frac{d\alpha d\beta}{8\pi^2}\,
\delta(1-x-\beta-\alpha+\lambda),
\end{eqnarray*}
we find that the partonic differential cross-section for a single gluon emission~is
\begin{eqnarray}
\label{DY_cs}
\frac{1}{\sigma}\frac{d\sigma}{dx}=
\frac{C_F \alpha_s}{2\pi}
\,\int_{\beta_1}^{\beta_2}
\,d\beta\,\left[
\frac{-\lambda}{(1-x-\beta)^2}\,(1-\beta)+\frac{1}{1-x-\beta}\,\frac{\beta^2-2\beta+2}{\beta}
\right].
\end{eqnarray}
The integration over $\beta$ should now be performed for fixed
$m^2$ and $s=2p\bar{p}$, thus for a fixed $\lambda$. As in the
previous cases, the relevant integration limit is deduced from the
condition that the gluon transverse momentum vanishes, i.e.
$\alpha\beta=\lambda$. Substituting the relation
$\alpha=1-x-\beta+\lambda$, one obtains a quadratic equation for
$\beta$ whose solutions are
$\beta_{1,2}=(1-x+\lambda\pm\Delta)/2$, where $\Delta\equiv
\sqrt{(1-x)^2-2\lambda(1+x)+\lambda^2}$. Integrating~(\ref{DY_cs})
one obtains the following characteristic function: \beq \left.{\cal
F}(x,\lambda)\right\vert_{\log}\,=\,\frac{2}{1-x}\,\ln\frac{(1-x-\lambda+\Delta)(1-x+\lambda+\Delta)}
{(1-x-\lambda-\Delta)(1-x+\lambda-\Delta)}, 
\label{F_DY}
\eeq where, as in
(\ref{F_frag}), a factor of $C_F\alpha_s/(2\pi)$ was extracted.
Since in this case only the small $\beta$ region contributes to log-enhanced terms, 
the only relevant term in (\ref{DY_cs}) is  
\begin{eqnarray}
\label{DY_cs1}
\frac{1}{\sigma}\frac{d\sigma}{dx}\simeq
\frac{C_F \alpha_s}{2\pi}
\,\int_{\beta_1}^{\beta_2}
\,d\beta\,\left[
\frac{2}{\left(1-x-\beta\right)\beta}
\right].
\end{eqnarray}
This simple expression is sufficient for the calculation of~(\ref{F_DY}).

Taking the logarithmic derivative with respect to $\lambda$ yields~\cite{BB_DY} 
\beq \left.{\dot {\cal
F}}(x,\lambda)\right\vert_{\log}\,=\,\frac{4}{\sqrt{(1-x)^2-2\lambda(1+x)+\lambda^2}}.
\label{dF_DY} \eeq
Comparing this characteristic function to the case of $e^+e^-$ annihilation (\ref{dF_frag}) 
or (\ref{dF_jets}) 
and  deep-inelastic scattering (\ref{dF_frag_DIS}), it is clear that the structure of the 
perturbative expansion and the power corrections will be completely different. 
This should be contrasted with the picture one might get based on the NLL resummation formula: in the
so-called deep inelastic factorization scheme~\cite{CMW}, the latter is identical to that of the thrust. 
In fact, the similarity of the leading logs is superficial. There is a fundamental difference between the first
set of examples, where the radiation originates in an {\em outgoing} quark (an evolving quark jet) 
with a constrained {\em invariant mass}, and the Drell--Yan case, where the radiation originates 
in the {\em incoming} quarks and the total {\em energy} ($\alpha+\beta$) in the final state is restricted. 
As we shall see this difference is encoded into the structure of the exponent making an impact on
both the sub-leading Sudakov logs and the power corrections.

We would like to express the SDG cross-section in the form
of a Borel sum, as in~(\ref{Borel_rep}). We should therefore
calculate  $\int {d\lambda}\,\lambda^{-u-1}{\dot {\cal
F}}(x,\lambda)$. As in the cases of the fragmentation functions
and deep inelastic structure functions, the lower limit is zero so
the integral diverges. Factorization must be used to obtain a
well-defined Borel function. Using the same subtraction
prescription as in~(\ref{Cq}) we get
\begin{eqnarray}
\label{B_SDG_DY}
B_{\SDG}(x,u) &=&
\,\frac{4}{1-x}\,\int_{1}^{(1-\sqrt{x})^2} \,d\lambda\,\lambda^{-u-1} \,\\
& &+\,\int_{0}^{(1-\sqrt{x})^2} {d\lambda} \,\lambda^{-u-1} \,
\,\left[\frac{4}{\sqrt{(1-x)^2-2\lambda(1+x)+\lambda^2}}-\frac{4}{1-x}
\right]. \nonumber
\end{eqnarray}
The upper integration limit was deduced
from the condition $\beta_{2}\leq \beta_1$, yielding~\hbox{$\Delta({\lambda_{\max}})=0$}, and therefore
$\lambda_{\max}=(1-\sqrt{x})^2$.
Since the integration over $\lambda$ is restricted to $\lambda \ll 1-x$, in the logarithmic approximation
one can replace the integration limit by
$\lambda_{\max}\simeq (1-x)^2/4$ and approximate the $\lambda$-dependent denominator in~(\ref{B_SDG_DY}) by
$\sqrt{(1-x)^2-4\lambda}$. We thus find
\begin{eqnarray}
\label{B_x_DY}
B_{\SDG}(x,u) &=&\frac{4}{1-x}\, \frac1u\,\left[
1-\,\frac{\sqrt{\pi}\,\Gamma(1-u)}{\Gamma(\frac12
-u)}\,\left(\frac{1-x}{2}\right)^{-2u}\right].
\end{eqnarray}
Finally, the Borel function corresponding to the exponent in
Mellin space is \beq \label{BN_DY} B_N(u)=\int_0^1 dx
\left(x^{N-1}-1\right)B_{\SDG}(x,u)=2\left(e^{2u\ln
N}-1\right)\Gamma(-u)^2-\frac4u\ln N. \eeq As far as the leading
terms at large $N$ are concerned, this result agrees with that of
ref.~\cite{BB_DY}, which used the full matrix element.

{\bf Sudakov logs:}
Equation~(\ref{BN_DY}) allows us to examine the structure of the exponent to a fixed logarithmic accuracy, as
in~(\ref{log_expansion}). Here $f_k(\xi)$ are given by
\beq
f_k(\xi)=\sum_{l=0}^{(k-1)/2}\,\frac{(-\pi^2)^l}{(2l+1)!}\,g_k^{(u)}(2\xi)\,M_{\rm in}(k,l)
\label{f_k_DY}
\eeq
where $g_k^{(l)}(\xi)$ is defined in~(\ref{g_kl}) and
\beq
M_{\rm in}(k,l)=2\sum_{j=-1}^{k-2l-2} \,c_{k-2l-3-j} \,c_j.
\eeq
Explicitly, this yields
\begin{eqnarray}
\label{f_k_12_DY}
f_{1}(\xi)&=& 2\,(1-2\xi)\,\ln(1 - 2\,\xi ) + 4\,\xi    \nonumber \\
f_{2}(\xi)&=& - 4\,\gamma \,\ln(1 - 2\,\xi )
\end{eqnarray}
and
\begin{eqnarray*}
\begin{array}{ll}
f_{3}(\xi)={1.33}/{(1-2\,\xi )} +6.58\,\xi \\
f_{4}(\xi)={2.12}/{(1-2\,\xi)^{2}}   \\
f_{5}(\xi)={4.00}/{(1 -2\,\xi)^{3}} - 19.48\,\xi  \\
f_{6}(\xi)={11.59}/{(1-2\,\xi)^{4}}  \\
f_{7}(\xi)={48.42}/{(1-2\,\xi)^{5}} +91.56\,\xi  \\
f_{8}(\xi)={238.80}/{(1-2\,\xi)^{6}}  \\
f_{9}(\xi)={1438.66}/{(1-2\,\xi)^{7}} - 527.14\,\xi  \\
f_{10}(\xi)={10078.0}/{(1-2\,\xi)^{8}}.
\end{array}
\end{eqnarray*}
As in the previous examples, sub-leading logs are enhanced by factorially increasing coefficients as well as an
increasing singularity at $\xi=\xi_{\max}=1/2$.

{\bf Power corrections:}
The power corrections in the $x\longrightarrow 1$ region can be deduced from the singularity of the
Borel transform of the exponent (\ref{BN_DY}). As in the previous examples, they are expected to exponentiate
together with the perturbative sum.
The singularities of  the Borel integrand are located at {\em integer} values of $u$. Equation~(\ref{BN_DY})
has double poles at all integers, but the $\sin(\pi u)/\pi u$ factor of~(\ref{Borel_A_eff})
leaves only simple poles. As observed in~\cite{BB_DY}, this singularity structure implies that the
leading power correction at large $Q^2$
(and not too large $x$) is $1/Q^2(1-x)^2$. Closer to $x=1$ sub-leading power corrections of the
form $1/(Q^2(1-x)^2)^n$, where $n$ is an integer, become important. The non-perturbative correction factor (in
Mellin space) is 
\beq
C_{\rm NP}(N,Q^2)=\exp\left\{\sum_{n=1}^{\infty}\frac{C_F\,(-1)^n}{{\beta_0}\,n\, (n!)^2}\,\omega_n
\,\left(\frac{{\bar{\Lambda}}^2N^2}{Q^2}\right)^n\right\}.
\eeq

\section{Conclusions}

The factorization of soft and collinear gluon bremsstrahlung is
one of the fundamental tools of perturbative QCD~\cite{Basics}. In
particular, it sets the basis for resummation of logarithms in
many applications. In this paper we demonstrated that
factorization holds also in the case of an off-shell gluon. The
main advantage in keeping the gluon off-shell is that the
virtuality provides the right physical scale for the coupling.
This opens the possibility to resum a full set of radiative
corrections, which are associated with dressing the emitted gluon (a
renormalon sum). Keeping the gluon off-shell, we identified the
appropriate kinematic approximation needed for the calculation of
log-enhanced terms originating in either the collinear or the soft
regions of phase space. The factorization formula led to the
definition of a generalized splitting function that describes the
emission probability of an off-shell gluon off a quark. As always,
factorization results in exponentiation. Thus, DGE promotes the
leading order calculation to an all-order sum in {\em two}
respects: the first by dressing the gluon and the second by taking
into account multiple emission through exponentiation.

In the class of problems analysed here, where a kinematic
threshold imposes a stringent restriction on the real emission
phase space, resummation is necessary even for a {\em qualitative}
description of the differential cross-section. Currently available
and future experimental data provide a challenge to QCD: {\em
quantitative} predictions for differential cross-section are
required. The threshold region turns out to be phenomenologically important in
many cases: for example in $e^+e^-$ annihilation the cross-section
is large only in the two-jet region.

The analysis of the exponent in the different examples considered
(see also \cite{DGE}) shows that sub-leading Sudakov
logs are enhanced factorially with respect to the leading logs. This
is a direct consequence of the integration over the running
coupling, which reveals the presence of infrared renormalons. Thus
because of the running coupling, a quantitative description of the
threshold region, e.g. in the fragmentation function case
$Q^2(1-x)\gsim \bar{\Lambda}$ or $(\beta_0\alpha_s/\pi)\,L \lsim
1$ (where $L=\ln 1/(1-x)$), requires the calculation of the
exponent to any logarithmic accuracy. Resummation with a fixed
logarithmic accuracy, e.g. NLL, applies only for
$(\beta_0\alpha_s/\pi)\,L \ll 1$. Moreover, in the threshold
region, non-perturbative corrections are particularly significant. As
opposed to single-scale observables, non-perturbative corrections
are not dominated by a leading power, but rather appear as a
convolution of the perturbative distribution with some {\em shape
function}~\cite{Shape_function1}--\cite{DGE} of the relevant
scale. We saw that the exponent calculated
through DGE contains both perturbative and non-perturbative 
information. As in the case of single-scale 
observables~\cite{Average_thrust}, 
these two aspects cannot be separated and must be treated 
together by resummation and parametrization of power corrections. 

As any renormalon-based approach, DGE does not pretend to supply a 
field theoretic definition of the shape function or to determine 
the magnitude of the non-perturbative corrections. However, it strongly suggests: 
$(1)$ the factorization of the non-perturbative corrections; $(2)$ 
the scale on which the non-perturbative function depends, e.g. $(1-x)Q^2$ in the fragmentation and 
deep inelastic cases and $Q(1-T)$ and $Q(1-x)$ in the thrust and Drell--Yan cases, respectively;
and $(3)$ the specific powers of this scale appearing in the exponent (in moment space), for example:
{\em odd} powers of $\nu/Q$ in the thrust, $(N/Q^2)^n$, where $n=1$ and $2$, in both the fragmentation and 
deep inelastic cases, and {\em even} powers of $N/Q$ in the Drell--Yan case.  
The first two properties can be deduced also from other considerations. The third is 
harder to access non-perturbatively. Here one really relies on the assumption 
that the dominant non-perturbative effects are the ones detected by 
the perturbative tools. 

\vskip 35pt
\noindent
{\bf Acknowledgements:} I wish to thank Yuri Dokshitzer, Gregory Korchemsky, Johan Rathsman, Gavin Salam
and Douglas Ross for very interesting and helpful discussions.


\begin{thebibliography}{9} 

\bibitem{Basics}
Y.~L.~Dokshitzer, V.~A.~Khoze, A.~H.~Mueller and S.~I.~Troian,
``Basics of perturbative QCD,'' (in ``Basics of'' series, 
Ed. Fronti${\rm \grave{e}}$res, Gif-sur-Yvette, 1991), 274 p.

\bibitem{Sterman:1987aj}
G.~Sterman,
{\it Nucl.\ Phys.}  {\bf B281} (1987) 310.

\bibitem{Collins:1988ig}
J.~C.~Collins, D.~E.~Soper and G.~Sterman,
{\it Nucl. Phys.}  {\bf B308} (1988) 833.

\bibitem{CT_DY}
S.~Catani and L.~Trentadue,
{\it Nucl. Phys.}  {\bf B327} (1989) 323.

\bibitem{Contopanagos:1997nh}
H.~Contopanagos, E.~Laenen and G.~Sterman,
{\it Nucl. Phys.} {\bf B484} (1997) 303 [hep-ph/9604313].

\bibitem{Low} F.E~Low, {\em Phys. Rev} {\bf 110}, (1958) 974 .

\bibitem{Burnett:1968km}
T.~H.~Burnett and N.~M.~Kroll,
{\it Phys. Rev. Lett.}  {\bf 20} (1968) 86.

\bibitem{Gribov} V.N. Gribov, ``Bremsshtralung of hadrons at high energies'',
{\it Sov. J. Nucl. Phys.}, {\bf 5} (1967) 280, {\it Yad. Fiz.}, {\bf 5} (1967) 399. 

\bibitem{DelDuca:1990gz}
V.~Del Duca,
{\it Nucl. Phys.}  {\bf B345} (1990) 369.

\bibitem{Chaichian:1995kq}
M.~Chaichian and B.~Ermolaev,
{\it Nucl. Phys.}  {\bf B451} (1995) 194.

\bibitem{BLM} S.J. Brodsky, G.P. Lepage and P.B. Mackenzie,
{\em Phys. Rev.} {\bf D28} (1983) 228; G.P.~Lepage and 
P.B.~Mackenzie, {\em Phys. Rev.} {\bf D48} (1993) 2250.

\bibitem{Disentangling}
S.~J.~Brodsky, E.~Gardi, G.~Grunberg and J.~Rathsman,
{\em Phys. Rev.}  {\bf D63} (2001) 094017 [hep-ph/0002065].

\bibitem{Average_thrust}
E.~Gardi and G.~Grunberg,
{\em JHEP} {\bf 9911} (1999) 016
[hep-ph/9908458].

\bibitem{beneke}
M.~Beneke,
{\em Phys. Rep.} {\bf 317} (1999) 1 [hep-ph/9807443].

\bibitem{Mueller}  A.H. Mueller, {\em Nucl. Phys.} {\bf B250}  (1985) 327;
{\em Phys. Lett. } {\bf B308} (1993) 355.

\bibitem{Zakharov} V.I. Zakharov, {\em Nucl. Phys.} {\bf B385} (1992) 452.

\bibitem{Shape_function1} G.P. Korchemsky and G. Sterman, {\em
Nucl. Phys.} {\bf B437} (1995) 415.

\bibitem{Shape_function2} G.P. Korchemsky and G. Sterman, [hep-ph/9505391]
 30th Rencontres de Moriond, {\em
QCD and high energy hadronic interactions}, Les Arcs, 1995. 
ed. J. Tran Thanh Van (Editions Fronti$\rm \grave{e}$res, Gif-sur-Yvette, 1995), p.383. 

\bibitem{Shape_function3}
G.~P.~Korchemsky,
{\em Shape functions and power-corrections to the event shapes},
hep-ph/9806537.

\bibitem{Shape_function4}
G.~P.~Korchemsky and G.~Sterman,
{\em Nucl. Phys.}  {\bf B555} (1999) 335 [hep-ph/9902341].

\bibitem{Korchemsky_Tafat}
G.~P.~Korchemsky and S.~Tafat,
{\em JHEP} {\bf 0010} (2000) 010 [hep-ph/0007005].

\bibitem{Belitsky:2001ij}
A.~V.~Belitsky, G.~P.~Korchemsky and G.~Sterman,
hep-ph/0106308.


\bibitem{DGE}
E.~Gardi and J.~Rathsman,
{\it Nucl. Phys.}  {\bf B609} (2001) 123 [hep-ph/0103217].

\bibitem{CTTW}
S.~Catani, L.~Trentadue, G.~Turnock and B.~R.~Webber,
{\it Nucl. Phys.}  {\bf B407} (1993) 3.

\bibitem{CS_DY}
H.~Contopanagos and G.~Sterman,
{\it Nucl. Phys.}  {\bf B419} (1994) 77 [hep-ph/9310313].

\bibitem{BB_DY}
M.~Beneke and V.~M.~Braun,
{\it Nucl. Phys.}  {\bf B454} (1995) 253 [hep-ph/9506452].

\bibitem{AZ_KLN}
R.~Akhoury, M.~G.~Sotiropoulos and V.~I.~Zakharov,
{\em Phys. Rev.} {\bf D56} (1997) 377
[hep-ph/9702270].

\bibitem{BB}
M.~Beneke and V.~M.~Braun,
{\em Phys. Lett.}  {\bf B348} (1995) 513 [hep-ph/9411229];
P.~Ball, M.~Beneke and V.~M.~Braun,
{\it Nucl. Phys.}  {\bf B452} (1995) 563
[hep-ph/9502300].

\bibitem{DMW}
 Yu.L. Dokshitzer, G. Marchesini and B.R. Webber,
{\em Nucl. Phys.} {\bf B469} (1996) 93.

\bibitem{CMW} S. Catani, G. Marchesini and B.R. Webber, {\em Nucl. Phys.}
 {\bf B349} (1991) 635.

\bibitem{CC}
M.~Cacciari and S.~Catani,
hep-ph/0107138.

\bibitem{DasW_frag}
M.~Dasgupta and B.~R.~Webber,
{\it Nucl. Phys.}  {\bf B484} (1997) 247 [hep-ph/9608394].

\bibitem{BBM}
M.~Beneke, V.~M.~Braun and L.~Magnea,
{\it Nucl. Phys.}  {\bf B497} (1997) 297
[hep-ph/9701309].

\bibitem{Korchemsky:1989si}
G.~P.~Korchemsky,
Mod.\ {\em Phys. Lett.}  {\bf A4} (1989) 1257.

\bibitem{Korchemsky:1993xv}
G.~P.~Korchemsky and G.~Marchesini,
{\it Nucl. Phys.}  {\bf B406} (1993) 225
[hep-ph/9210281];
{\em Phys. Lett.}  {\bf B313} (1993) 433.

\bibitem{BY}  L.S. Brown and L.G. Yaffe, {\em Phys. Rev.} {\bf D45} (1992)
 398;  L.S. Brown, L.G. Yaffe and C. Zhai, {\em Phys. Rev.} {\bf D46}
(1992) 4712.

\bibitem{B2} M. Beneke, {\em Nucl. Phys.} {\bf B405} (1993) 424.

\bibitem{G3} G. Grunberg, {\em Phys. Lett.} {\bf B304} (1993) 183;
see also {\em Quantum Field Theoretic Aspects of High Energy Physics},
Kyffhauser, Germany, September 1993.

\bibitem{wise}
A.~V.~Manohar and M.~B.~Wise,
{\em Phys. Lett.} {\bf B344} (1995) 407
[hep-ph/9406392].

\bibitem{W}
B.~R.~Webber,
{\em Phys. Lett.}  {\bf B339} (1994) 148 [hep-ph/9408222].

\bibitem{DW} Yu.L. Dokshitzer and B.R. Webber,  {\em Phys. Lett.}
{\bf B352} (1995) 451 and 
{\bf B404} (1997) 321.

\bibitem{AZ}
R.~Akhoury and V.~I.~Zakharov,
{\em Phys. Lett.}  {\bf B357} (1995) 646 [hep-ph/9504248].

\bibitem{higher_moments}
E.~Gardi,
{\em JHEP} {\bf 0004} (2000) 030 [hep-ph/0003179].

\bibitem{Heavy_Jet}
E.~Gardi and J.~Rathsman, ``Sudakov logs 
and renormalons in the thrust and the jet-mass distributions'', 
{\em to be published}.

\bibitem{EFP}
R.~K.~Ellis, W.~Furmanski and R.~Petronzio,
{\it Nucl. Phys.}  {\bf B207} (1982) 1;
{\it Nucl. Phys.}  {\bf B212} (1983) 29.

\bibitem{JS}
R.~L.~Jaffe and M.~Soldate,
{\em Phys. Rev.} {\bf D26} (1982) 49.

\bibitem{DasW_DIS}
M.~Dasgupta and B.~R.~Webber,
{\em Phys. Lett.}  {\bf B382} (1996) 273 [hep-ph/9604388].

\bibitem{Stein:1996wk}
E.~Stein, M.~Meyer-Hermann, L.~Mankiewicz and A.~Schafer,
{\em Phys. Lett.}  {\bf B376} (1996) 177
[hep-ph/9601356].

\bibitem{Meyer-Hermann:1996cy}
M.~Meyer-Hermann, M.~Maul, L.~Mankiewicz, E.~Stein and A.~Schafer,
{\em Phys. Lett.}  {\bf B383} (1996) 463
[Erratum-ibid.\  {\bf B393} (1996) 487]
[hep-ph/9605229].

\bibitem{Maul:1997rz}
M.~Maul, E.~Stein, A.~Schafer and L.~Mankiewicz,
{\em Phys. Lett.}  {\bf B401} (1997) 100
[hep-ph/9612300].

\bibitem{Akhoury:1997rt}
R.~Akhoury and V.~I.~Zakharov,
hep-ph/9701378.

\bibitem{DIS}E.~Gardi, G.P.~Korchemsky, D.A.~Ross and S.~Tafat, 
``Power corrections in deep inelastic scattering at large Bjorken x'', {\em to be published}.

\bibitem{Vogt}
A.~Vogt,
{\em Phys. Lett.}  {\bf B497} (2001) 228 
[hep-ph/0010146].

\end{thebibliography}
\end{document}